\documentclass[twocol]{ametsoc}

\journal{jas}

\bibpunct{(}{)}{;}{a}{}{,}

\usepackage{cuted}
\usepackage{caption}
\title{Convective dynamics and the response of precipitation extremes to warming in radiative-convective equilibrium}

\authors{Tristan H. Abbott\correspondingauthor{Department of Earth, Atmospheric, and Planetary Sciences, MIT, Cambridge, Massachusetts.}}

\affiliation{Department of Earth, Atmospheric, and Planetary Sciences, MIT, Cambridge, Massachusetts}

\email{thabbott@mit.edu}

\extraauthor{Timothy W. Cronin}
\extraaffil{Department of Earth, Atmospheric, and Planetary Sciences, MIT, Cambridge, Massachusetts}

\extraauthor{Tom Beucler}
\extraaffil{Department of Earth System Science, University of California, Irvine, California\\Department of Earth and Environmental Engineering, Columbia University, New York, New York}

\abstract{Tropical precipitation extremes are expected to strengthen with warming, but quantitative estimates remain uncertain because of a poor understanding of changes in convective dynamics. This uncertainty is addressed here by analyzing idealized convection-permitting simulations of radiative-convective equilibrium in long-channel geometry. Across a wide range of climates, the thermodynamic contribution to changes in instantaneous precipitation extremes follows near-surface moisture, and the dynamic contribution is positive and small, but sensitive to domain size. The shapes of mass flux profiles associated with precipitation extremes are determined by conditional sampling that favors strong vertical motion at levels where the vertical saturation specific humidity gradient is large, and mass flux profiles collapse to a common shape across climates when plotted in a moisture-based vertical coordinate. The collapse, robust to changes in microphysics and turbulence schemes, implies a thermodynamic contribution that scales with near-surface moisture despite substantial convergence aloft and allows the dynamic contribution to be defined by the pressure velocity at a single level. Linking the simplified dynamic mode to vertical velocities from entraining plume models reveals that the small dynamic mode in channel simulations ($\lesssim$2\% K$^{-1}$) is caused by opposing height-dependences of vertical velocity and density, together with the buffering influence of cloud-base buoyancies that vary little with surface temperature. These results reinforce an emerging picture of the response of extreme tropical precipitation rates to warming: a thermodynamic mode of about 7\% K$^{-1}$ dominates, with a minor contribution from changes in dynamics.}

\begin{document}

\maketitle

%








\section{Introduction}

Tropical precipitation extremes are expected to intensify under global warming \citep[e.g.][]{IPCC2013ClimateChange,Allen2002ConstraintsCycle,Pendergrass2014ChangesWarming}. To first order, the intensification is driven by changes in the thermodynamic structure of the atmosphere. If moisture is converged into updrafts primarily near the surface, then approximate expressions for condensation rates suggest that---absent significant changes in convective dynamics---peak precipitation rates should scale with boundary layer saturation specific humidity and increase with surface temperatures at about 7 \% K$^{-1}$ \citep[e.g.][]{OGorman2009TheChange,Muller2011IntensificationModel}. However, the baseline 7 \% K$^{-1}$ sensitivity of peak precipitation rates to warming can be modified by convective dynamics. If convective storms converge moisture throughout an inflow layer that extends several kilometers above the surface---as has been observed in Amazonian convection \citep{Schiro2018GoAmazon2014/5Scales}---precipitation rates may scale with lower-troposphere-average water vapor, which increases with warming faster than boundary layer water vapor. Furthermore, changes in updraft speeds affect condensation rates and thus precipitation extremes \citep[e.g.][]{Emori2005DynamicClimate,OGorman2009TheChange,Muller2011IntensificationModel,Romps2011ResponseWarming,Pendergrass2016TheSimulations}, introducing a dynamic contribution to changes in precipitation extremes in addition to the thermodynamic dependence. Given the potentially severe social and economic consequences of extreme rain events, understanding the influence of convective dynamics on changes in precipitation extremes in a warming climate is crucial for our ability to prepare for the impacts of future hydrologic change.

A common scaling for assessing the impact of dynamic and thermodynamic changes on precipitation extremes equates a high-percentile precipitation rate $P$ to the product of a column-integrated condensation rate $C$ and a precipitation efficiency $\epsilon$ \citep{OGorman2009TheChange,Muller2011IntensificationModel}:
\begin{equation}
    P = \epsilon C.
\end{equation}
Depending on the time scale considered, $C$ may be spatiotemporally co-located with extreme precipitation events \citep[as in][]{Muller2011IntensificationModel,Fildier2017SimultaneousChange} or may be a high percentile of the column-integrated condensation rate distribution \citep[as in][]{Singh2014TheTemperature}. The condensation rate in an adiabatically lifted parcel can then be used to re-write column condensation in terms of pressure-coordinate vertical velocity $\omega$ and saturation specific humidity $q^*$ \citep{OGorman2009TheChange}:
\begin{equation} \label{eqn:condint}
    C = -\int_{0}^{p_s} \frac{\omega}{g} \left(\frac{\mathrm{d}q^*}{\mathrm{d}p} \right)_{\theta_e^*} \mathrm{d}{p}.
\end{equation}
In Equation \ref{eqn:condint}, $p$ is pressure, $p_s$ is surface pressure, $g$ is the acceleration from gravity, and the change in saturation specific humidity with pressure $\left({\rm d}q^*/{\rm d}p\right)_{\theta_e^*}$ (hereafter referred to as the ``moisture lapse rate'') is calculated following a moist adiabat (i.e., holding saturated equivalent potential temperature $\theta_e^*$ fixed). 

The scaling can then be used to relate the response of precipitation extremes to warming to changes in precipitation efficiency, changes in the moisture lapse rate (the ``thermodynamic mode''), and changes in pressure velocity profiles (the ``dynamic mode'') \citep{Muller2011IntensificationModel}:
\begin{strip}
\begin{equation}
    \frac{\delta P}{P} \approx \frac{\delta \epsilon}{\epsilon} + \underbrace{\frac{-\int_{0}^{p_s} \frac{\omega}{g} \delta \left(\frac{\mathrm{d}q^*}{\mathrm{d}p} \right)_{\theta_e^*} \mathrm{d}{p}}{C}}_\text{Thermodynamic mode} + \underbrace{\frac{-\int_{0}^{p_s} \delta \left ( \frac{\omega}{g} \right) \left(\frac{\mathrm{d}q^*}{\mathrm{d}p} \right)_{\theta_e^*} \mathrm{d}{p}}{C}}_\text{Dynamic mode}.
\label{eqn:decomp}
\end{equation}
\end{strip}
Here, $\delta$ indicates a change between two climates, and terms that include products of changes have been dropped. Throughout this paper, we use the terms "thermodynamic mode" and "dynamic mode" to refer to contributions to changes in precipitation extremes corresponding to the two terms labeled in Eq. \ref{eqn:decomp}.

Previous modeling studies have found, empirically, that the thermodynamic mode at temperatures close to Earth's present-day tropics is close (within about 2 \% K$^{-1}$) to the $\approx$7 \% K$^{-1}$  Clausius-Clapeyron (CC) scaling of saturation specific humidity with surface temperature, regardless of whether the studies use global simulations with parameterized convection \citep{OGorman2009TheChange,Ogorman2009ScalingGCM,Chen2019ThermodynamicEvents}, global simulations with superparameterized convection \citep{Fildier2017SimultaneousChange}, or limited-area radiative-convective equilibrium (RCE) simulations with convection-permitting models \citep{Muller2011IntensificationModel,Romps2011ResponseWarming,Muller2013ImpactWarming}. A CC scaling for the thermodynamic mode can also be derived theoretically by making three assumptions: that (1) $\omega/g$ is constant between the surface and the tropopause, (2) $q^*$ is near-zero at the tropopause, and (3) the temperature profile is moist adiabatic \citep{Ogorman2009ScalingGCM,Muller2011IntensificationModel}. If all three assumptions hold, the thermodynamic mode reduces approximately to a CC scaling of $\delta(q^*_s)/q^*_s$, where $q^*_s$ is the near-surface saturation specific humidity. 

In the tropics, where the tropopause is cold and convection maintains nearly moist adiabatic stratification \citep{Xu1989IsUnstable}, assumptions 2 and 3 are clearly justified, but assumption 1 is not. On both large scales and convective scales, vertical motion profiles associated with heavy tropical rainfall are far from constant with height \citep{Back2006GeographicPacific,Torri2017StableTropics,Moore2014AEffect}. Moreover, although the empirical success of the CC scaling might suggest that the thermodynamic mode is insensitive to the shape of the mass flux profiles, simple calculations with synthetic mass flux profiles indicate otherwise. In moist adiabatic atmospheres, moisture lapse rates increase with warming most rapidly high in the atmosphere, particularly in warm climates (Figure \ref{fig:f1}a; see Appendix A for a discussion of the upper-tropospheric amplification). As a result, column-integrated condensation produced by top-heavy vertical motion profiles increases with warming more quickly than condensation produced by bottom-heavy mass flux profiles. For top-heavy vertical motion profiles, the thermodynamic mode can be nearly 11 \% K$^{-1}$ in cold climates, and for bottom-heavy vertical motion profiles, the thermodynamic mode can be as low as 3\% K$^{-1}$ in warm climates (Figure \ref{fig:f1}b,c). The sensitivity of the thermodynamic mode to the shape of mass flux profiles prompts the first question addressed in this paper: What constrains the shape of pressure velocity profiles associated with precipitation extremes, and are the constraints consistent with a near-CC thermodynamic mode across a wide range of climates?

Empirical results from previous work using convection-permitting models also suggest that the dynamic mode is likely to be small. On both hourly and daily timescales, the dynamic mode in idealized convection-permitting model simulations rarely exceeds $\sim$2\% K$^{-1}$, and the sign varies between studies \citep{Muller2011IntensificationModel,Romps2011ResponseWarming,Muller2013ImpactWarming}. Global simulations with the superparameterized Community Atmosphere Model (SP-CAM) also produce a small ($\le 2$\% K$^{-1}$) dynamic mode for precipitation extremes aggregated across the tropics \citep{Fildier2017SimultaneousChange}. We lack a similar empirical consensus from global simulations with parameterized convection, but we believe the wide spread of dynamic modes in individual models---which largely owes to large inter-model differences in vertical velocities \citep{Ogorman2009ScalingGCM,OGorman2009TheChange,Norris2019ThermodynamicEnsemble,Pendergrass2014ChangesWarming}---should be treated with skepticism. Twentieth century simulations show large differences in precipitation extremes between different models \citep{Kharin2007ChangesSimulations}, and many models contain disproportionate increases in extreme precipitation rates relative to the rest of the precipitation distribution because of increases in precipitation linked to artificial gridpoint storms \citep{Pendergrass2014ChangesWarming,Pendergrass2014TwoRain}.

Although convection-permitting models provide little reason to expect a large dynamic mode, there is no simple scaling that predicts a small dynamic mode. Moreover, zero-buoyancy plume models constrained by the observation that the tropical stratification is set by strongly entraining updrafts \citep{KuangBretherton2006,Romps2009DoTroposphere,Singh2013InfluenceEquilibrium} predict strong increases in convective available potential energy (CAPE) and peak vertical velocities with warming \citep{Singh2015IncreasesEquilibrium,Seeley2015WhyWarming,Romps2016ClausiusClapeyronRCE}.
\citet{Muller2011IntensificationModel} point out that the dynamic  mode for changes in precipitation extremes can be negative despite increases in peak vertical velocities if vertical motion weakens near the surface, and \citet{Fildier2017SimultaneousChange} argue that near-CC increases in CAPE with warming \citep{Seeley2015WhyWarming,Romps2016ClausiusClapeyronRCE} provide a 1/2-CC upper bound on the dynamic mode. Nevertheless, the relationship between increases in CAPE, peak vertical velocities, and precipitation extremes remains unclear, prompting the second question addressed in this paper: Can a better understanding of controls on the shapes of vertical motion profiles associated with precipitation extremes provide insight into the relationship between changes in peak updraft velocities and the dynamic mode?

In this paper, we address these questions by analyzing instantaneous precipitation extremes in idealized RCE simulations with a convection-permitting model in a 12,000 km non-rotating channel over sea surface temperatures (SSTs) between 280 and 305 K. Section \ref{sec:lcs} documents the simulations, shows that they produce precipitation rate probability density functions (PDFs) similar to PDFs calculated from optical rain gauges in the tropical western Pacific, and defines the extreme events analyzed in this paper. Thereafter, the paper is organized into three main sections. In Section \ref{sec:scal}, we decompose changes in precipitation extremes with SST into contributions from changes in precipitation efficiency, thermodynamic modes, and dynamic modes. Consistent with previous work, we find a near-CC thermodynamic mode and a small but positive dynamic mode. We also find that contributions from changes in precipitation efficiency are comparable in magnitude to the dynamic mode, although this is not the focus of the paper. A novel result is that pressure velocity profile shapes associated with high-percentile column-integrated condensation collapse across climates when plotted in a moisture-based vertical coordinate. We show that this collapse implies a near-CC thermodynamic mode, and use the collapse to derive a simplified expression for the dynamic mode. In Section \ref{sec:pprof}, we combine the simplified expression for the dynamic mode with an entraining plume model to explain why the dynamic mode is small. In Section \ref{sec:mod}, we use a simplified version of the entraining plume model to derive a theoretical model for the dynamical mode that is valid so long as the pressure velocity profile shape collapse holds. This model predicts a dynamic mode that is about 1 \% K$^{-1}$ at temperatures close to the modern day tropics and decreases, albeit weakly, with increasing SST. We conclude in Section \ref{sec:conc} by discussing remaining questions and directions for future work.

\section{Long-Channel Simulations of Radiative-Convective Equilibrium} \label{sec:lcs}
\subsection{Simulation Setup}

The long-channel RCE simulations analyzed in this paper use the System for Atmospheric Modeling (SAM) version 6.8.2 \citep{Khairoutdinov2003CloudSensitivities} in an elongated 12,288 km $\times$ 192 km channel with 3 km horizontal resolution. The lower boundary is a sea surface with uniform fixed temperature, varied between simulations from 280 K to 305 K  in increments of 5 K. Each simulation is 75 days long, and large-scale overturning circulations that emerge during convective self-aggregation are fully developed after day 50. Both the humidity contrast between moist and dry regions and the length scale of overturning circulations vary between simulations \citep{Wing2016Self-aggregationGeometry,Beucler2019ASelf-aggregation}, but neither vary strongly or monotonically with SST, making it difficult to clearly link changes in the degree of aggregation with changes in precipitation extremes. The simulations are documented in more detail in \citet{Wing2016Self-aggregationGeometry}, with an emphasis on the processes that drive convection self-aggregation, and \citet{Cronin2017CloudsModel} describe clouds, large-scale circulations, and climate sensitivity in a similar set of channel simulations. \citet{Wing2016Self-aggregationGeometry} include an additional 310 K SST simulation in their analysis, but we choose to exclude that simulation from this paper because, without increasing the height of the model top, convection impinges on the bottom of a sponge layer at the top of the model domain.

The analysis in this paper focuses on instantaneous rather than hourly or daily precipitation extremes. The reasons for this decision are twofold. First, previous studies on changes in CAPE and updraft velocities with warming have focused on instantaneous updraft velocities, and analyzing the dynamic mode for instantaneous precipitation extremes provides a closer link with that work. Second, archived output from many convection-permitting model simulations (including our channel simulations) only includes instantaneous snapshots of the three-dimensional fields required to diagnose thermodynamic and dynamic modes, and focusing on instantaneous precipitation allows us to re-purpose data from past modeling studies. The instantaneous three-dimensional snapshots on which we base our analyses are output from the channel simulations once every six hours. We analyze precipitation extremes over the last 25 days of each simulation, a time period during which convection is aggregated into moist and dry bands and all simulations are in a statistically steady state.

We have chosen to focus our analysis on channel simulations rather than general circulation model (GCM) simulations with parameterized convection or small-domain simulations of unaggregated RCE for two reasons. First, our channel simulations explicitly simulate motions from the convective scale (O(1 km)) to the planetary scale (O(10,000 km)) and can produce convective precipitation extremes without relying on parameterizations of convection---unlike GCM simulations. Second, our channel simulations produce convective systems organized at scales similar to the real tropics and to global cloud-resolving models \citep{Cronin2017CloudsModel,Beucler2019ComparingEquator}---unlike small-domain RCE. While convective organization makes the channel simulations more complex than small-domain RCE, the lack of convective organization means that small-domain RCE simulations are unable to represent the organized convective systems that produce a large fraction of real-world tropical rainfall \citep[e.g.][]{Nesbitt2006StormFeatures}.

In Appendix B, we briefly document some important similarities and differences between convective dynamics in channel versus small-domain RCE. One conspicuous difference is that the dynamic mode is often smaller in small-domain RCE than in channel simulations (subsection c of Appendix B). Given this discrepancy, we should emphasize that the results in this paper are specific to a particular class of RCE simulations (i.e., RCE in an elongated channel), albeit one that we feel is of particular relevance for the study of precipitation extremes.

\subsection{Defining Instantaneous Precipitation Extremes}

As in previous work, we define precipitation extremes based on quantiles of the precipitation rate distribution. In long-channel RCE simulations, instantaneous precipitation rate PDFs have a delta function at zero rain rate, decay like a power law at low rain rates, and roll off into an exponential tail at high rain rates. This structure is consistent with observed PDFs \citep{Martinez-Villalobos2019WhyDistributions}: the instantaneous rain rate PDF from our 305 K SST simulation is very similar to a one-minute-average rain rate PDF calculated from optical rain gauge measurements at Nauru Island (in the Tropical Western Pacific, with surrounding SSTs typically between 300 to 305 K) (Figure \ref{fig:f2}). 99.9th percentile instantaneous precipitation rates\footnote{Our definition of percentiles includes ``events'' with rain rates of zero, so that the rain rate $r(p)$ at a percentile $p$ is defined by
\begin{equation*}
\int_{0}^{r(p)} P(r') {\rm d}r' = p,
\end{equation*}
where $P(r)$ is the rain rate PDF and includes the delta function at $r = 0$.} lie on the edge of the power law range, which means that the likelihood of observing a particular precipitation rate falls relatively slowly as the rain rate increases above the 99.9th percentile. On the other hand, 99.99th percentile instantaneous precipitation rates fall firmly within the exponential tail of the rain rate PDF, which means that the likelihood of observing a particular rain rate drops very quickly above the 99.99th percentile. A closer examination of events that produce rain near the 99.9th and 99.99th percentiles reveals significant differences in underlying dynamics: while 99.99th percentile instantaneous precipitation rates almost always occur within deep convective systems, 99.9th percentile instantaneous precipitation rates often occur in relatively shallow clouds that terminate far below the tropopause. In other words, 99.99th percentile instantaneous precipitation is associated with statistically and dynamically extreme events in a way that 99.9th percentile precipitation is not. Because our simulation output contains many events at and above the 99.99th percentile (over 2500 total, and on average 26 per snapshot), allowing analysis with relatively little sample error, we focus the remainder of this paper on 99.99th precipitation rates.

The choice to focus on a specific quantile warrants some discussion of the potential quantile-dependence of changes in precipitation extremes. Previous studies disagree on whether changes in precipitation extremes converge with increasing quantile, although convergence generally seems more common in simulations with resolved convection (e.g. \citet{Muller2011IntensificationModel}, \citet{Muller2013ImpactWarming}, and the SPCAM simulations in \citet{Fildier2017SimultaneousChange}) and less common in simulations with parameterized convection (see e.g. \citet{Pendergrass2014ChangesWarming} and the CAM simulations in \citet{Fildier2017SimultaneousChange}). Our simulations, which model convection explicitly, are consistent with this trend in that they show convergence at high quantiles: precipitation extremes increase more quickly with warming at the 99.99th percentile than at the 99th or 99.9th percentiles, but the rate of change is similar between the 99.99th and 99.999th percentiles (Figure \ref{fig:f2}b). Decomposing these changes into dynamic and thermodynamic modes (not shown) reveals that the dynamic mode also converges with warming to typical values of around +2 \% K$^{-1}$ at the 99.99th percentile and above, although we do not analyze the processes that control the dynamic mode in detail for percentiles other than the 99.99th.

\section{Scaling of Simulated Precipitation Extremes} \label{sec:scal}

Because the time lag between condensation and precipitation fallout prevents us from applying the condensation integral scaling directly to columns with 99.99th percentile surface precipitation, we instead apply the scaling to columns where the column condensation rate is near the 99.99th percentile. Although 99.99th percentile column condensation rates are higher than 99.99th percentile precipitation rates by about a factor of 2 (indicating a typical precipitation efficiency for extreme events of $\epsilon \approx 0.5$), both increase with warming at close to a CC scaling (Figure \ref{fig:f3}a). When calculating condensation integrals, we set pressure velocities to zero in regions where the atmosphere is unsaturated. (This has little impact on our analysis of 99.99th percentile condensation extremes, as these typically occur in columns where the air is saturated over the entire depth of the troposphere.) We calculate the pressure velocity profile $\omega$ used to decompose changes in condensation integrals into dynamic and thermodynamic modes as an average over the 1000 columns with condensation rates closest to the 99.99th percentile column condensation rate, and we calculate the the moisture lapse rate $\left({\rm d}q^*/{\rm d}p\right)_{\theta_e^*}$ based on time- and domain-average temperature profiles over the last 25 days of simulation. When calculating the moisture lapse rate, we follow the saturation vapor pressure formulation from the SAM single-moment microphysics scheme \citep{Khairoutdinov2003CloudSensitivities} (which the channel simulations use) and scale saturation vapor pressure linearly from the vapor pressure over liquid at 273.16 K to the vapor pressure over ice at 253.16 K. This makes the temperature derivative of saturation vapor pressure discontinuous at 273.16 K and 253.16 K, which introduces small discontinuities in the moisture lapse rate that appear as minor kinks when the profiles are plotted. Choosing to calculate moisture lapse rates based on local rather than mean temperature profiles has a negligible impact on our results.

The decomposition into dynamic and thermodynamic modes is easiest to visualize by plotting rates of change as a function of surface temperature (Figure \ref{fig:f3}b). Doing so reveals that both condensation and precipitation rates increase with warming at a rate close to (but, for column condensation, consistently higher than) a CC scaling. Changes in precipitation efficiency between climates vary in sign, and a simple estimate of sampling uncertainty\footnote{We estimate the sampling uncertainty for changes with warming by bootstrapping. Specifically, we repeat our calculations 1000 times after resampling individual snapshots with replacement and take the minimum and maximum rates of change from the 1000 sets of calculations as the minimum and maximum of our uncertainty intervals.} suggests that much of the variation may be a result of sampling error. Unlike \citet{Singh2014TheTemperature}, we find no evidence of a strong microphysical influence on changes in precipitation extremes in relatively cold climates, likely because the SAM single-moment microphysics scheme we use \citep{Khairoutdinov2003CloudSensitivities} produces fairly slow increases in hydrometeor fall speed with warming \citep{Lutsko2018IncreaseEquilibrium}. The thermodynamic mode follows a CC scaling nearly exactly over the entire range of simulated climates; the dynamic mode is positive, generally smaller than the thermodynamic mode, and largest in cold climates.

Rather than defining the decomposition using pressure velocity and pressure coordinates, we could instead have defined a dynamic mode in height coordinates ($z$) based on height-coordinate vertical velocity ($w$) by re-writing the condensation integral as
\begin{equation}
    C = - \int_{0}^{\infty} \rho w \left (\frac{{\rm d}q^*}{{\rm d}z} \right )_{\theta_e^*} {\rm d}z
\end{equation}
and decomposing changes as
\begin{equation} \label{eqn:zdecomp}
\begin{split}
	\delta C = &-\int_{0}^{\infty} \delta w \left ( \rho \left (\frac{{\rm d}q^*}{{\rm d}z} \right )_{\theta_e^*} \right ) {\rm d}z \\
	&- \int_{0}^{\infty} w \delta \left ( \rho \left (\frac{{\rm d}q^*}{{\rm d}z} \right )_{\theta_e^*} \right ) {\rm d}z.
\end{split}
\end{equation}
The choice to calculate changes in profiles at fixed pressure (as is implied by Equation \ref{eqn:decomp}) versus fixed height (as is implied by Equation \ref{eqn:zdecomp}) has a negligible impact on the decomposition (pressures at fixed heights change by not more than 15 hPa between adjacent simulations). The choice to calculate the dynamic mode in terms of changes in $\omega$ verus $w$, on the other hand, obviously affects the partitioning of changes between dynamic and thermodynamic modes. We chose a decomposition that separates $\omega$ and $\left({\rm d}q^*/{\rm d}p\right)_{\theta_e^*}$ primarily because it expresses the thermodynamic mode in terms of changes in the moisture lapse rate alone, which is consistent with and allows for easy comparison with previous studies \citep[e.g.][]{Ogorman2009ScalingGCM,OGorman2009TheChange,Muller2011IntensificationModel,Muller2013ImpactWarming, Fildier2017SimultaneousChange}.

Our results (Figure \ref{fig:f3}) support the robustness of a CC scaling for the thermodynamic mode, but exactly why the CC scaling is so robust remains unclear. In the following section, we examine pressure velocity and moisture lapse rate profiles to understand how changes in the vertical structure of extreme events combine to produce a near-CC thermodynamic mode over a wide range of climates.

\subsection{Role of Vertical Structure}

Pressure velocity profiles associated with high condensation rates are far from constant with height (Figure \ref{fig:f4}a), so the thermodynamic mode could in theory differ considerably from CC-scaling (as in Figure \ref{fig:f1}).  Unlike the profiles used to create Figure \ref{fig:f1}, however, the simulated pressure velocity profiles shift upward as SST increases, with a bottom-heavy vertical structure in cold climates and a more top-heavy vertical structure in warm climates (Figure \ref{fig:f4}a). The upward shift with warming is crucial for maintaining a near-CC thermodynamic mode: thermodynamic modes calculated across the entire SST range using the pressure velocity profile from the coldest simulation (280 K) are sub-CC in warm climates; likewise, thermodynamic modes calculated using the pressure velocity profile from the warmest simulation (305 K) are super-CC in cold climates (Figure \ref{fig:f4}c).

Because the radiative tropopause moves upward with warming in RCE \citep{Seeley2019FATInvariant}, deep convection occupies a deeper layer of the atmosphere in warmer simulations, and this alone could contribute to an upward shift in the level of peak pressure velocities with warming. Additionally, though, moisture lapse rates are bottom-heavy in cold climates and become increasingly top-heavy with warming (Figure \ref{fig:f4}b). Changes in the vertical structure of the moisture lapse rate could also drive an upward shift in pressure velocity profiles when those profiles are conditionally sampled from columns with large column-integrated condensation rates.

We can probe the impact of the shapes of moisture lapse rate profiles on the shapes of conditionally-sampled pressure velocity profiles by asking the following question: what pressure velocity profile shapes would we find in columns with 99.99th percentile condensation {\it if the moisture lapse rate were constant with height}? This is equivalent to calculating profiles of $\omega$ conditionally averaged over columns with near-99.99th percentile column-integrated mass flux,
\begin{equation}
    M = -\int_{0}^{p_s} \frac{\omega}{g} {\rm d}p,
\end{equation}
because the true condensation integral $C$ depends only on $\omega$ and moisture lapse rate profiles (recall Equation \ref{eqn:condint}). If similar $\omega$ profiles emerged when conditionally averaging on high-percentile condensation $C$ and high-percentile mass flux $M$, this would indicate that vertical variations of the moisture lapse rate do not constrain $\omega$ profiles in columns with high-percentile condensation. If different $\omega$ profiles emerged when conditioning on high-percentile $C$ compared to $M$, however, this would indicate that vertical structure in the moisture lapse rate acts to constrain $\omega$ profiles that lead to $C$ extremes. In fact, we find that pressure velocity profiles conditioned on extreme $M$ shift upward less dramatically with warming than those conditioned on extreme $C$ (compare Figures \ref{fig:f5}a and \ref{fig:f5}b). This indicates that the vertical structure of the moisture lapse rate {\it does} play an important role in driving the upward shift of $\omega$ profiles with warming, and therefore must influence the shapes of $\omega$ profiles in strongly-condensing columns.

But how exactly does the vertical structure of the moisture lapse rate influence the shape of pressure velocity profiles when conditionally averaged on strong column condensation? The key is that the (explicitly-simulated) convective dynamics in our simulations produce a diversity of individual pressure velocity profiles---some very bottom-heavy, some very top-heavy, and some in between. Although many different profile shapes can produce large convective mass fluxes, large condensation rates are only produced when mass fluxes are large at levels of the atmosphere where the moisture lapse rate is also large. In cold climates, on the one hand, the moisture lapse rate is only large close to the surface, so only pressure velocity profiles with strong vertical motion near the surface are capable of producing large condensation rates. As a result, pressure velocity profiles conditionally averaged on high-percentile column condensation are bottom-heavy. In warm climates, on the other hand, the moisture lapse rate remains large much farther from the surface, so somewhat more top-heavy pressure velocity profiles often produce large column-integrated condensation rates. As a result, pressure velocity profiles conditionally averaged on extreme column condensation shift upward relative to cold climates.

An important component of the argument in the previous paragraph is the fact that the deepest and strongest vertical motion profiles do not necessarily produce the strongest condensation rates. This result is reminiscent of the observation by \citet{Hamada2015WeakStorms} that TRMM radar reflectivity profiles are typically much more bottom-heavy for pixels with the heaviest surface rainfall than for pixels with the highest 40-dBZ echo top heights typically associated with extremely strong convection.

\subsection{Collapse in a moisture-based vertical coordinate}

A succinct way to summarize the link between moisture lapse rate and pressure velocity profile shapes is to say that strongly condensing pressure velocity profile shapes are determined in large part by solving an optimization problem: given a moisture lapse rate, what pressure velocity shapes are best able to produce large column-integrated condensation rates? In cold climates, these shapes tend to be bottom-heavy; in warm climates, less so. Because we have shown in Section 3a that the moisture lapse rate is the dominant parameter in this optimization problem, we expect the solution---that is, the pressure velocity profile shape---to depend largely on the moisture lapse rate itself. This motivates looking at the shapes of pressure velocity profiles in a vertical coordinate based on the moisture lapse rate.

A vertical coordinate defined as a normalized integral of the moisture lapse rate,
\begin{equation}
    \tilde{q}(p) = \frac{ \int_{0}^{p} \left (\frac{\rm{d}q^*}{{\rm d}p} \right)_{\theta_e^*} {\rm d}p' }{ \int_{0}^{p_s} \left (\frac{\rm{d}q^*}{{\rm d}p} \right)_{\theta_e^*} {\rm d}p'},
\end{equation}
is a convenient choice for four reasons. First, it has a relatively simple interpretation: if temperature profiles are moist adiabatic, then $\left(\mathrm{d}q^*/\mathrm{d}p \right)_{\theta_e^*}$ (the rate at which saturation specific humidity changes under adiabatic expansion) is equal to the vertical gradient of saturation specific humidity $\partial q^*/\partial p$ and $\tilde{q}$ simplifies to $q^*/q^*_s$, i.e. saturation specific humidity normalized by surface saturation specific humidity. Second, $\tilde{q}$ has a climate-invariant range: it is always equal to 1 at the surface and decreases to a minimum value of 0 within the troposphere. Third, the use of $\tilde{q}$ simplifies the condensation integral, which can be expressed as a $\tilde{q}$-coordinate integral of pressure velocity alone rather than a pressure-coordinate integral of the product of pressure velocity and the moisture lapse rate. Fourth, the moisture lapse rate is close to its maximum value over a similar range of $\tilde{q}$ coordinates across a wide range of climates (Figure \ref{fig:f5}c). Because $\omega$ profiles in strongly condensing columns tend to be large at levels where the moisture lapse rate is also large, this suggests that the shapes of $\omega$ profiles may vary less across climates in $\tilde{q}$ coordinates than in pressure coordinates. Indeed, if we plot the shapes of pressure velocity profiles as functions of $\tilde{q}$, they nearly collapse on top of each other (Figure \ref{fig:f5}d).

We should emphasize that although we expect $\tilde{q}$ to be a convenient vertical coordinate, our finding that $\omega$ profile are nearly climate-invariant in $\tilde{q}$ coordinates is largely an empirical rather than a theoretical result. Nonetheless, the collapse appears to be reasonably robust across a range of model configurations; it also occurs in small-domain RCE simulations with a range of microphysics and sub-grid-scale turbulence schemes (see Appendix B, subsection b).  Additionally, the collapse dramatically simplifies both the dynamic and thermodynamics modes, and we leverage these simplifications throughout the rest of this paper.

In particular, it is possible to show that an exact collapse of $\omega$ in $\tilde{q}$ coordinates implies an exactly CC thermodynamic mode if the temperature profile is moist adiabatic. For moist adiabatic temperature profiles, $\left(\mathrm{d}q^*/\mathrm{d}p \right)_{\theta_e^*}$ is equal to the vertical gradient of saturation specific humidity $\partial q^*/\partial p$. The assumption of moist adiabatic temperature profiles allows us to rewrite the condensation integral in terms of $q^*$:
\begin{equation}
\begin{split}
C &= -\int_{0}^{p_s} \frac{\omega}{g} \left(\frac{{\rm d} q^*}{{\rm d}p}\right)_{\theta_e^*} {\rm d}p \\ &= -\int_{0}^{p_s} \frac{\omega}{g} \frac{\partial q^*}{\partial p} {\rm d}p = -\int_{0}^{q^*_s} \frac{\omega}{g} {\rm d}q^*,
\end{split}
\end{equation}
and to simplify $\tilde{q}$ to $\tilde{q} = q^*/q^*_s$. If the collapse is exact, then $\omega = \hat{\omega} \Omega(\tilde{q})$, where $\hat{\omega}$ is the maximum pressure velocity attained along the profile and $\Omega(\tilde{q})$ is a universal function across climates. The condensation scaling can then be simplified to
\begin{equation}
C = - \hat{\omega} q^*_s  g^{-1} \gamma,
\end{equation}
where
\begin{equation}
    \gamma = \int_{0}^{1} \Omega(\tilde{q}) {\rm d}\tilde{q}
\end{equation}
is a unitless parameter that depends on the shape of pressure velocity profiles. Because an exact collapse implies that $\gamma$ is constant across climates, changes in the condensation scaling can be written as
\begin{equation} \label{eqn:sdecomp}
\frac{\delta C}{C} = \frac{\delta \hat{\omega}}{\hat{\omega}} + \frac{\delta q^*_s}{q^*_s}.
\end{equation}
This gives a thermodynamic mode that scales exactly with changes in surface saturation specific humidity and a dynamic mode that depends on changes in the pressure velocity at a single level rather than on moisture-lapse-rate-weighted changes in entire pressure velocity profiles. \citet{Fildier2017SimultaneousChange} also decompose the dynamic mode into a contribution from a characteristic pressure velocity and a residual contribution from changes in vertical structure. However, they use a characteristic pressure velocity based on column-integrated pressure velocity profiles; unlike our decomposition, this results in a residual term that is comparable in magnitude to the overall dynamic mode.

We end this section by providing some physical interpretation of our results. Past derivations of a CC scaling \citep[e.g.][]{OGorman2009TheChange,Muller2011IntensificationModel} have frequently relied on the assumption that strong updrafts converge moisture near the surface and diverge moisture only at levels where the saturation specific humidity is near-zero. The pressure velocity profiles implied by this assumption would have $\Omega(\tilde{q})$ = 0 below the LCL ($\tilde{q} = \textrm{ surface relative humidity}$) and $\Omega(\tilde{q}) = 1$ above it, and therefore have $\gamma$ equal to surface relative humidity. For peaked mass flux profiles, on the other hand, $\gamma$ is strictly smaller than surface relative humidity ($\gamma \approx 0.5$ in our simulations) and can be thought of as a convergence-related efficiency factor: it is reduced either by converging moisture as levels where $q^* < q^*_s$ or by diverging moisture at levels where $q^*/q^*_s$ is non-negligible. Thus, we obtain a near-CC scaling not because strong updrafts consistently condense most near-surface water vapor, but rather because profiles of convergence and divergence reduce condensation by the same factor over a wide range of climates. This reconciles the success of the CC scaling with previous work that has found that convective precipitation is frequently associated with significant moisture convergence aloft \citep{Moore2014AEffect}.

In the next section, we use a simple entraining plume model to understand changes in $\hat{\omega}$ between climates; given the collapse in $\omega$ profiles this amounts to a model for the dynamic mode.

\section{Controls on Peak Pressure Velocity} \label{sec:pprof}

The collapse of pressure velocity profile shapes in a moisture-based vertical coordinate allows us to write the dynamical mode in terms of changes in the pressure velocity at the peak of the pressure velocity profile. In turn, changes in the peak pressure velocity can be decomposed into changes in density $\rho$ and height-coordinate vertical velocity $w$:
\begin{equation*}
    \frac{\delta \hat{\omega}}{\hat{\omega}} = \frac{\delta \rho}{\rho} + \frac{\delta w}{w}.
\end{equation*}
In the channel simulations, changes in height coordinate vertical velocities contribute between 0 \%/K and +5 \%/K to the dynamic mode, while changes in density contribute between 0 \%/K and -3 \%/K (Figure \ref{fig:f6}). The dynamic mode remains small (between about 0\%/K and +2 \%/K) because increases in vertical velocities are consistently offset by decreases in density.

The reductions in density at the level of peak pressure velocity can largely be understood in terms of changes in density at fixed $\tilde{q}$ in a moist adiabatic atmosphere. As the surface warms, levels with fixed $\tilde{q}$ shift to lower pressures and density at fixed $\tilde{q}$ drops as a result (Figure \ref{fig:f7}a). At levels where $\tilde{q} \approx 0.3$ to $0.4$ (where conditional-average pressure velocity profiles peak), density decreases with surface temperature at rates between 1 \%/K and 3 \%/K (Figure \ref{fig:f7}b), roughly consistent with the changes in density shown in Figure \ref{fig:f6}.

The increases in height-coordinate vertical velocities can be explained in terms of changes in buoyancy integrals produced by an entraining plume model \citep[based on][]{Singh2015IncreasesEquilibrium}. To do so, we assume that the vertical velocity $w$ of an ascending parcel at a physical height $z$ scales with the integral of the buoyancy $b$ experienced by the parcel during ascent from cloud-base height $z_b$ such that \citep[e.g.][]{Holton2013AnMeteorology}
\begin{equation}
    w(z) = \sqrt{2 \eta \int_{z_b}^{z} b(z') \mathrm{d} z'}.
\end{equation}
$\eta$ is an empirically-chosen proportionality factor that we set to $0.14$ for a good fit to simulated vertical velocities. The relatively coarse horizontal resolution of our simulations may be part of the reason that a good fit is provided by $\eta$ much smaller than $1$: in updrafts that are wide relative to their height, non-hydrostatic pressure gradients cancel a substantial part of parcel buoyancy \citep{Jeevanjee2016EffectiveAloft}. Form drag, which is an important component of the momentum budget of thermals \citep{Romps2015StereoThermals,Jeevanjee2015EffectivePools,Morrison2018TheoreticalThermals} may also help to keep $\eta$ small.

We calculate buoyancy profiles by integrating equations for liquid/ice water static energy $h_L$ and moisture $q$ within an entraining plume:
\begin{align}
    \frac{\mathrm{d} h_L}{\mathrm{d}z} &= - \epsilon (h_L - h_{Le}) \\
    \frac{\mathrm{d} q}{\mathrm{d}z} &= - \epsilon (q - q_e).
\end{align}
Liquid/ice water static energy is a moist-static-energy-like variable used as a prognostic thermodynamic variable by SAM and is defined as $h_L = c_p T + g z - L_v q_l - L_f q_i$ \citep{Khairoutdinov2003CloudSensitivities}, where $c_p$ is the heat capacity at constant pressure of dry air, $T$ is temperature, $L_v$ is the latent heat of vaporization, $L_f$ is the latent heat of fusion, $q_l$ is the mass concentration of liquid water, and $q_i$ is the mass concentration of ice water. $q$ is the total water mass concentration (including vapor, liquid, and ice phases); our plume model assumes no condensate fallout. We define the environmental energy profile $h_{Le}$ based on the time- and domain-mean temperature profile $T_e$ from the simulations so that $h_{Le} = c_p T_e + gz$, and we define the environmental water profile based on the time and domain-mean saturation specific humidity profile $q^*_e$ and a constant environmental relative humidity of 90 \%. When combined with information about saturation specific humidity as a function of pressure and temperature, $h_L$ and $q$ can be inverted for plume temperature $T$, specific humidity $q_v$, and condensate mass concentration $q_l + q_i$. We then calculate buoyancy following its definition in SAM as $b = g \left ( (T - T_e)/T_e + 0.608 (q_v - q_e) - (q_l + q_i) \right )$. We initialize the plume at a cloud-base height of $z_b = 1$ km by finding an initial plume $h_L$ and $q$ that gives an in-plume relative humidity of 100 percent with no condensed water and produces a buoyancy equal to the 99.99th percentile buoyancy at the model level closest to 1 km. We integrate the plume equations upward using an entrainment rate of $\epsilon = 0.15 \textrm{ km}^{-1}$, chosen for a good overall fit to simulated buoyancy profiles and buoyancy integrals.

Buoyancy integrals calculated from entraining plume model buoyancy profiles have similar magnitude to integrals based on level-by-level 99.99th percentile buoyancies\footnote{The 99.99th percentile level-by-level buoyancies are calculated from all samples taken at each level. However, we set buoyancies to zero unless they are co-located with ascending cloudy air in an attempt to separate large buoyancies in convective updrafts from large buoyancies in gravity waves.} (Figure \ref{fig:f8}a). When integrated from cloud base to the height where conditional-average pressure velocity profiles peak, both versions of the buoyancy integral increase with warming, from about 70 J kg$^{-1}$ in the coldest simulation to about 275 J kg$^{-1}$ in the warmest. Accordingly, vertical velocities predicted based on the plume model buoyancy integrals increase with warming, and if we choose $\eta = 0.14$, they agree well with the height-coordinate vertical velocities in conditional-average pressure velocity profiles at the levels where the pressure velocity profiles peak (Figure \ref{fig:f8}b). The plume model vertical velocities also provide good predictions for the 99.99th percentile vertical velocity taken over all samples from the level where conditional-average pressure velocity profiles peak.

Examining buoyancy profiles in more detail can provide some qualitative understanding of the general smallness of the dynamic contribution to changes in precipitation extremes. As noted by \citet{Singh2015IncreasesEquilibrium}, buoyancy in weakly entraining plumes increases with warming primarily in the upper troposphere. In our simulations and in our plume model calculations, warming produces only minor changes in buoyancy at fixed heights below about 5 km (Figure \ref{fig:f9}a,b)---a feature of RCE that is discussed in detail by \citet{Seeley2015WhyWarming}. Note that peaks in simulated buoyancy above the above the tropopause in Figure \ref{fig:f9}a are not associated with convective plumes that originate in the troposphere, and so are not reproduced by the plume model in Figure \ref{fig:f9}b. Because bottom-heavy moisture lapse rate profiles only allow bottom-heavy pressure velocity profiles to produce large column-integrated condensation rates, conditional-average pressure velocity profiles peak at or below 5 km from the surface in all simulations (recall e.g. Figure \ref{fig:f4}). As a result, increases in buoyancy integrals and height-coordinate vertical velocities come primarily from an upward shift in the heights at which pressure velocity profiles peak. However, this upward shift also results in decreases in density at the level of peak pressure velocity. As a result, the contributions to peak conditional-average pressure velocities from changes in height-coordinate vertical velocities and densities offset to produce a consistently small dynamic mode.

The buoyancy profiles produced by the plume model also emphasize the importance of the cloud-base buoyancy. The diagnosed buoyancy is on the order of 0.05 m s$^{-2}$ across all SSTs (Figure \ref{fig:f9}c), and alternative plume model integrations with zero cloud-base buoyancy significantly underestimate lower-tropospheric buoyancy compared to simulations (Figure \ref{fig:f9}d).

\section{Theoretical Model for Changes in Precipitation Extremes} \label{sec:mod}

In Section \ref{sec:scal}, we showed that simulated pressure velocity profiles in columns with high-percentile column condensation rates collapse to a climate-invariant shape when plotted in $\tilde{q}$-coordinates. In Section \ref{sec:pprof}, we provided a simple explanation for the magnitude of the dynamic mode by combining calculations of density along moist adiabatic temperature profiles with vertical velocities calculated with an entraining plume model. In this section, we build on those two results by constructing a simple, self-contained model for the dynamic mode. This model assumes that the $\tilde{q}$-coordinate collapse holds in all climates (more specifically, that conditional-average pressure velocity profiles peak at $\tilde{q} = 0.35$) and calculates peak pressure velocities using a simplified entraining plume model based on \citet{Singh2013InfluenceEquilibrium} and \citet{Singh2015IncreasesEquilibrium}. The reasons for constructing this model are two-fold. First, it provides a prediction for the dynamic mode (and thus for changes in condensation extremes themselves) for a wider range of climates than the simulations alone. Second, and potentially more importantly, it provides a way to probe the processes that control the dynamic mode within the simulated range of climates without the sampling errors introduced by analyzing a finite-duration simulation, without the full complexity of chaotic convective dynamics, and without noisiness introduced by non-monotonic changes in large-scale organization with warming \citep{Wing2016Self-aggregationGeometry}.

\subsection{Derivation}

The starting point for the model is Equation 10 from \citet{Singh2015IncreasesEquilibrium}, which relates the temperature anomaly $\Delta T_w$ in a weakly entraining plume to the plume entrainment rate and environmental parameters:
\begin{equation}
\begin{split}
	\Delta T_w &= \frac{1}{1 + \frac{L_v^2 q^*_e(z)}{c_p R_v T_e^2(z)}} \bigg [ e^{-\varepsilon_w \left ( z - z_b \right )} \frac{\Delta h_b}{c_p} \\
	&+ \frac{\Delta \varepsilon \left(1 - RH \right) L_v}{c_p}\int_{z_b}^{z} e^{-\varepsilon_w (z - z')} q^*_e(z) \mathrm{d}z' \bigg ].
\end{split}
\end{equation}
This equation is modified slightly compared to Equation 10 of \citet{Singh2015IncreasesEquilibrium} to include a cloud-base moist static energy (MSE) anomaly $\Delta h_b$. $\varepsilon_w$ is the entrainment rate of the weakly entraining plume, $\Delta \varepsilon$ is the difference in entrainment rate between the weakly entraining plume and the more common strongly-entraining plumes that set environmental temperature profiles, $q^*_e(z)$ is the environmental saturation specific humidity, $RH$ is the environmental relative humidity, $R_v$ is the specific gas constant for water vapor, $T_e(z)$ is the environmental temperature profile, and $z_b$ is the cloud-base height. 

To calculate environmental profiles $T_e(z)$ and $q^*_e(z)$, we use a zero-buoyancy plume model following \citet{Singh2013InfluenceEquilibrium}. This model integrates
\begin{equation*}
    \frac{{\rm d}h}{{\rm d}z} = -\varepsilon_s L_v \left(1 - RH\right) q^*_e,
\end{equation*}
where $h$ is the plume MSE and $\varepsilon_s = \varepsilon_w + \delta \varepsilon$ is the zero-buoyancy plume entrainment rate, from cloud base upward. At each vertical level, the plume temperature can be obtained from $h$ and the assumption that the plume is saturated, and the zero-buoyancy assumption then allows us to calculate $q^*_e$ at that level by assuming that the plume and environmental temperatures are equal. We initialize the zero-buoyancy plume model at a cloud-base height diagnosed assuming a surface relative humidity of 0.8, and we use $\varepsilon_s = 0.5$ km$^{-1}$.

Once $T_e$ is known, calculating buoyancy integrals akin to those used in Section \ref{sec:pprof} requires numerically evaluating two integrals: one to calculate $\Delta T_w$ as a function of height, and a second to calculate buoyancy integrals from $\Delta T_w(z)$. For this model, we neglect the impact of water vapor virtual effects and condensate loading and use $g \Delta T_w / T_e$ as an estimate of plume buoyancy. As in Section \ref{sec:pprof}, we use 1 km as a proxy for cloud-base height, and motivated by the relatively small variations in cloud-base buoyancy anomalies with SST found in Figure \ref{fig:f9}c, we diagnose cloud-base MSE anomalies that result in climate-invariant buoyancies at cloud base. We assume $RH = 0.9$ (again, as in Section \ref{sec:pprof}), and we repeat the calculation for several cloud-base buoyancies between 0 and 0.1 m s$^{-2}$ and several entrainment rates $\varepsilon_w$ between $0$ and $0.3$ km$^{-1}$. We set $\varepsilon_w$ to $0.15$ km$^{-1}$ (as in Section \ref{sec:pprof}) for calculations where we vary cloud-base buoyancy, and we set the cloud-base buoyancy to 0.05 m s$^{-2}$ (close to typical values from Figure \ref{fig:f9}c) for sets of calculations where we vary $\varepsilon_w$.

After calculating buoyancy integrals from cloud base up to the height where $\tilde{q} = 0.35$, we then assume that the height-coordinate vertical velocity at that level scales with the buoyancy integral as in Section \ref{sec:pprof}:
\begin{equation}
    w = \sqrt{2 \eta g \int_{z_b}^{z(\tilde{q} = 0.35)} \frac{\Delta T_w(z')}{T_e(z')} \mathrm{d}{z'}}
\end{equation}
with $\eta = 0.14$. Finally, we combine the vertical velocity with the density at $\tilde{q} = 0.35$ to calculate a pressure velocity $\omega = -\rho g w$. By repeating this calculation for many different surface temperatures, we can obtain values for the dynamic mode (i.e. $\mathrm{d} \ln{\omega}/\mathrm{d}SST$, assuming the shape collapse holds) over a wide range of climates.

\subsection{Results}

Both height-coordinate and pressure-coordinate vertical velocities increase with warming in the simple model, but offsetting changes in density mean that pressure velocity increases less quickly (Figure \ref{fig:f11}a,b). At low temperatures, the impact of density changes is relatively minor, and pressure-coordinate vertical velocities increase only slightly less quickly than height-coordinate vertical velocities. As temperatures increase, however, density decreases more and more rapidly at fixed $\tilde{q}$ (as in Figure \ref{fig:f7}b), and pressure velocities start to plateau at surface temperatures around 310 to 320 K. As a result, the dynamic mode decreases as surface temperature increases. The cloud-base buoyancy has a major impact on the vertical velocities and pressure velocities predicted by the simple model: larger cloud-base buoyancies produce stronger vertical motion, and a cloud-base buoyancy of 0.05 m s$^{-2}$ (similar to the values diagnosed from simulations in Figure \ref{fig:f9}c) provides a good fit to simulated vertical velocities. Cloud-base buoyancy also affects the dynamic mode, with larger cloud-base buoyancies producing a smaller dynamic mode, particularly in cold climates. This occurs largely because vertical velocities are very small in cold climates when cloud-base buoyancy is weak, and even small absolute increases in vertical velocity with warming translate into large relative increases and contribute to a large dynamic mode. As for vertical velocities and pressure velocities, the simulated dynamic mode is fit well by a cloud-base buoyancy anomaly of 0.05 m s$^{-2}$.

In constrast with the sensitivity to the cloud-base buoyancy, the dynamic mode in the simple model is relatively insensitive to the entrainment rate of the weakly entraining plume (Figure \ref{fig:f11}d). Larger entrainment rates produce slightly more negative dynamic modes, consistent with the lower rate of increase of vertical velocities with warming when entrainment is stronger \citep{Singh2015IncreasesEquilibrium}. However, entrainment rates between 0 and 0.3 km$^{-1}$ all provide reasonably good fits to the simulated dynamic mode. This suggests that the boundary layer processes that influence cloud-base buoyancy may play a more important role in setting the dynamic mode than the processes that control mixing between convective updrafts and their environment.

Because the shape collapse assumed by this simple model implies a near-CC thermodynamic mode, the temperature dependence of the dynamic mode implies a super-CC increase in condensation extremes with warming in cold climates and an approach to a near-CC or slightly sub-CC increase in very warm climates. Assumptions about cloud-base buoyancy and, to a lesser extent, entrainment have a quantitative impact on the rate at which the dynamic mode decreases as temperature increases, but the qualitative result that the dynamic mode decreases in very warm climates is robust to changes in both parameters. Whether the structural assumptions in the simple model remain valid for SSTs above 305 K is an open question, however, and the simple model could break down if the pressure velocity profile collapse fails to hold or the entraining plume calculation no longer provides a good model for peak pressure velocities.

\subsection{Discussion}

This simple model differs in important ways from the entraining-plume-based model for precipitation extremes from \citet{Loriaux2013UnderstandingModel}. In the model from \citet{Loriaux2013UnderstandingModel}, entire vertical velocity profiles are calculated by integrating an entraining plume model upward from the surface, with environmental profiles taken from mid-latitude observations rather than from a zero-buoyancy plume model. This approach provides a single vertical velocity profile, with vertical velocity profile shapes that are unaffected by conditional sampling, and produces a super-CC thermodynamic mode. Like our model, their model produces a dynamic mode of around 2 \%/K at surface temperatures around 300 K, although \citet{Loriaux2013UnderstandingModel} explore a much smaller temperature range ($\pm$ 3 K only). Because our simple model agrees well with RCE simulations while the entraining plume model of \citet{Loriaux2013UnderstandingModel} agrees well with midlatitude observations of convective precipitation extremes, a more detailed comparison of predictions from the two models may provide a way to probe differences between the response of tropical versus midlatitude convective precipitation to warming.

Finally, the dynamic mode predicted by this simple model can violate the 1/2-CC upper bound proposed by \citet{Fildier2017SimultaneousChange} based on the Clausius-Clapeyron scaling of CAPE with warming. In RCE, contributions to increases in CAPE with warming come largely from the upper troposphere \citep{Singh2013InfluenceEquilibrium,Seeley2015WhyWarming,Seeley2016TropicalIce} and so primarily affect upper-tropospheric vertical velocities. In constrast, the dynamic mode in our simple model is more closely linked to vertical velocities in the low-to-mid-troposphere and can increase more rapidly than CAPE with warming because changes in the upper limit of buoyancy integrals alter the fraction of CAPE accessed by high-condensation updrafts. Although opposing changes in vertical velocities and densities with height and the buffering effects of large cloud-base buoyancies generally keep the dynamic mode small, the dynamic mode can nevertheless exceed a 1/2-CC scaling with surface temperature in cold climates.

\section{Conclusion} \label{sec:conc}

In an effort to better understand how convective dynamics influence changes in tropical precipitation extremes in a warming world, we analyzed changes in 99.99th percentile instantaneous precipitation extremes in long-channel simulations of RCE across a wide range of climates. We focused on two primary questions: first, on how the shape of convective pressure velocity profiles mediates the response of precipitation extremes to atmospheric moisture content (the ``thermodynamic mode''); and second, on how precipitation extremes are affected by changes in pressure velocity profiles between climates (the ``dynamic mode''). Our analysis produced four primary results
\begin{enumerate}
    \item The shapes of pressure velocity profiles associated with precipitation extremes result from a conditional sampling of convective dynamics, and the conditional sampling favors profiles with large pressure velocities at levels in the atmosphere where the vertical saturation specific humidity gradient is large (Figure \ref{fig:f4}).
    \item Because of this conditional sampling, the shapes of pressure velocity profiles collapse across climates when plotted using a moisture-based vertical coordinate $\tilde{q} = q^*/q^*_s$ (Figure \ref{fig:f5}). Because of this collapse, the thermodynamic mode follows a Clausius-Clapeyron scaling with surface temperature across a wide range of climates, and the dynamic mode reduces to changes in profile-maximum pressure velocities.
    \item The simplification to the dynamic mode enabled by the $\tilde{q}$-coordinate collapse allows us to link the magnitude of the dynamic mode to changes in vertical velocities predicted by entraining plume models (Figures \ref{fig:f8}). This, in turn, allows us to argue that the smallness of the dynamic mode (compared to the thermodynamic mode) is linked to the insensitivity of lower-tropospheric convective buoyancy to warming (Figure \ref{fig:f9}) and to decreases in density as the level of peak conditional-average ascent shifts upward (Figure \ref{fig:f7}).
    \item Taking the $\tilde{q}$-coordinate collapse as a given allows us to develop a self-contained simple model for changes in condensation extremes. This simple model highlights that, in addition to opposing changes in vertical velocities and densities with height, climate-invariant cloud base buoyancies help to maintain a small dynamic mode. For a cloud-base buoyancy that provides a good fit to simulations, the simple model predicts a dynamic mode that is largest ($\sim$+2 \% K$^{-1}$) in climates much colder than today's tropics, small but positive ($\sim$+1 \%/K) at temperatures close to the present-day tropics, and weakest ($\sim$0 to -1 \% K$^{-1}$) in very warm climates (Figure \ref{fig:f11}).
\end{enumerate}

Given that many of these results rely on the $\tilde{q}$-coordinate collapse, for which we currently have only a qualitative explanation, some discussion of the robustness of our results is in order. One of the overarching themes of this paper is that $\tilde{q}$ coordinates are a natural vertical coordinate for thinking about hydrologic extremes, and that this coordinate allows for a potentially cleaner separation of dynamic and thermodynamic changes than more conventional vertical coordinates. In our case, viewing the vertical structure of instantaneous extremes in $\tilde{q}$ coordinates led to significant simplifications to the thermodynamic and dynamic modes because of a collapse of pressure velocity profile shapes in those coordinates. It is entirely possible that a similar collapse will fail to hold on other timescales, in other types of models, or for other types of precipitation. For example, ascent profiles in orographic precipitation extremes may be controlled primarily by topography and may therefore fail to shift upward with warming to the extent required for a $\tilde{q}$-coordinate collapse. Similarly, the collapse may fail to occur in models with convective parameterizations that place strong constraints on the vertical structure of updraft mass flux profiles. Nevertheless, viewing the dynamics of extreme events in a moisture-based vertical coordinate may provide useful insight into how the interplay between dynamics and thermodynamics affects the strength of precipitation extremes, and may suggest novel theoretical constraints to one or both of the thermodynamic and dynamic modes.

Even more generally, framing precipitation extremes as an optimization problem (``generate as much condensation as possible, subject to thermodynamic and dynamic constraints'') could be adapted to a diverse range of scenarios to better understand and constrain precipitation extremes in a changing climate. Doing so might appear to add an additional layer of complexity to an already difficult problem, but we found that it allowed us to connect the vertical structure of convective updrafts (which are difficult to constrain in general) to the thermodynamic structure of the tropical atmosphere (which, in contrast, is fairly well-understood). Although the dominant constraints in the optimization problem may vary in different situations, simply identifying what those constraints are may reveal hidden connections between precipitation extremes and their environment.

Our results also prompt a number of questions that could motivate future work. One obvious question is whether the predictions from the simple model for the dynamic mode presented in Section \ref{sec:mod} hold in RCE simulations of very cold or very warm climates. Additionally, the sensitivity of the simple model's dynamic mode to cloud-base buoyancy raises the question of what sets cloud-base buoyancy in extreme precipitation or condensation events in the real world---a subject that might be amenable to observational study. Another set of questions revolves around whether the shapes of vertical motion profiles associated with tropical precipitation extremes are as strongly constrained by moisture lapse rate profiles on longer timescales or larger spatial scales. If the moisture lapse rate constraint remains strong on larger spatial scales and longer time scales, does this mean that time- and space-averaged pressure velocity profiles continue to collapse in $\tilde{q}$ coordinates? Or, if the collapse does not hold on larger spatial scales and longer time scales, does this lead to significant deviations of the thermodynamic mode from a Clausius-Clapeyron scaling? Finally, observational data may provide an additional avenue for probing the robustness of the $\tilde{q}$-coordinate collapse. When combined with sufficiently accurate estimates of atmospheric temperature profiles, the convective vertical motion profiles retrieved by radar wind profilers provide the information required to compute condensation integrals from observational data. If observed vertical motion profiles associated with high-percentile condensation integrals also collapse across different temperatures in $\tilde{q}$ coordinates, this would provide evidence that the collapse is a robust emergent property of the dynamics that produce convective precipitation extremes.

Finally, we want to finish by discussing how we view this work in the broader context of research on the response of precipitation extremes to warming. In our view, the main contributions of this paper are (1) to show that the thermodynamic mode is near-CC and the dynamic mode is small in long-channel RCE simulations, consistent with the emerging consensus from simulations that explicitly resolve convection, and (2) to explain why this is the case {\it in our channel RCE simulations specifically}. Because the dynamic mode in the real tropics is likely to be affected by features (e.g. land-ocean contrasts, rotation, and topography) that the channel simulations lack, it is very unlikely that our results will generalize directly to every model configuration, and comprehensive simulations will play a leading role in improving constraints on future changes in precipitation extremes. Nevertheless, insights gained from idealized studies of precipitation extremes in RCE are valuable. Process-oriented analysis of idealized models improves our understanding of how the world works and, in doing so, influences how we view it. By highlighting several mechanisms that help to produce a near-CC thermodynamic mode and a small dynamic mode, this paper provides a foundation for using 7 \% K$^{-1}$ as a baseline estimate for the rate at which tropical precipitation extremes intensify with warming.

%

\acknowledgments
Tristan H. Abbott was supported by a Norman C. Rasmussen Fellowship and a Grayce B. Kerr Fellowship. Tristan H. Abbott and Timothy W. Cronin were supported by NSF grant AGS-1740533, and Tom Beucler was supported by NSF grants AGS-1520683 and OAC-1835769. The authors thank Marat Khairoutdinov for providing SAM, Allison Wing for providing simulation output, Nadir Jeevanjee and two anonymous reviewers for helpful comments, and Kathleen Schiro for providing the Nauru optical rain gauge time series. The Nauru rain gauge time series was collected by the U.S. Department of Energy Atmospheric Radiation Measurement Tropical West Pacific facility and is available at www.archive.arm.gov. Processed model output and code required to reproduce the figures in this paper is available at github.com/thabbott/JASPrecipitationExtremes, and raw SAM output is available from the authors upon request (email thabbott@mit.edu).

\clearpage

%






%
%
%

\appendix[A]
\appendixtitle{Shapes of moisture lapse rate profiles}
\label{app:moisture_lapse_rate}

The moisture lapse rate profiles shown in Figures \ref{fig:f1}a and \ref{fig:f4}b have three important features: first, they decline more slowly with height than either saturation vapor pressure or saturation specific humidity; second, they increase at fixed pressure with increasing surface temperature, and third, the increase is amplified in the upper troposphere in warm climates.

We can get some intuition for these features by writing down and analyzing an expression for the moisture lapse rate:
\begin{equation} \label{eqn:mlrexp}
\begin{split}
	\left(\frac{{\rm d}q^*}{{\rm d}p}\right)_{\theta_e^*} &= \frac{M_v}{M_a}\left(\frac{{\rm d}(e^*/p)}{{\rm d}p}\right)_{\theta_e^*} \\
	&= \frac{M_v}{M_a} \left ( \frac{1}{p} \left(\frac{{\rm d}e^*}{{\rm d}p}\right)_{\theta_e^*} - \frac{e^*}{p^2} \right ).
\end{split}
\end{equation}
where $M_v$ and $M_a$ are the molecular weights of water vapor and dry air and $e^*$ is the saturation vapor pressure.

We can use an approximate version of the Clausius-Clapeyron relation \citep[][]{Pierrehumbert2010PrinciplesClimate} to re-write the derivative of saturation vapor pressure as
\begin{equation} \label{eqn:clcl}
\begin{split}
\    	\left(\frac{{\rm d}e^*}{{\rm d}p}\right)_{\theta_e^*} &= \frac{L_v e^*}{R_v T^2} \left(\frac{{\rm d}T}{{\rm d}p}\right)_{\theta_e^*} \\
	&= \frac{L_v e^*}{R_v T^2} \frac{R_a T}{c_p p} \phi(T,p),
\end{split}
\end{equation}
where 
\begin{equation}
    \phi(T,p) = \frac{1 + \frac{L_v q^*}{R T}}{1 + q^* \left (\frac{c_{pv}}{c_p} + \frac{L_v}{c_p T}\left( \frac{L_v}{R_v T} - 1\right) \right)} \in [0,1]
\end{equation}
is the ratio between the moist and dry adiabatic temperature lapse rates \citep[e.g.][]{Pierrehumbert2010PrinciplesClimate} and $c_{pv}$ is the heat capacity at constant pressure of water vapor. Substituting Equation \ref{eqn:clcl} into Equation \ref{eqn:mlrexp} and re-arranging gives
\begin{equation} \label{eqn:mlr}
\begin{split}
	\left(\frac{{\rm d}q^*}{{\rm d}p}\right)_{\theta_e^*} &= \frac{M_v}{M_a} \left ( \frac{1}{p} \frac{L_v e^*}{R_v T^2} \frac{R_a T}{c_p p} \phi(T,p) - \frac{e^*}{p^2} \right ) \\
	&= \frac{M_v}{M_a} \frac{e^*}{p^2} \left ( \frac{T_0}{T} \phi(T,p) - 1 \right ),
\end{split}
\end{equation}
where $T_0 \approx 1550$ K is a temperature scale given by
\begin{equation}
T_0 = \frac{L_v R_a}{R_v c_p}.
\end{equation}

Equation \ref{eqn:mlr} shows that the moisture lapse rate is set by the product of two profiles: $M_v e^*/(M_a p^2)$, which mostly decreases with height but does so less quickly than saturation specific humidity (Figure A1a); and $(T_0/T)\phi - 1$, which always increases with height along moist adiabats (Figure A1b). The combined influence of these two profiles produces the three features noted at the start of this section:
\begin{enumerate}
    \item The moisture lapse rate decreases away from the surface much less quickly away than saturation vapor pressure. When the surface is warm and the water vapor scale height is sufficiently large, the moisture lapse rate can even increase with height.
    \item As the surface warms, $e^*/p^2$ increases while $(T_0/T)\phi - 1$ decreases. Although these changes offset, they combine to produce an increase in the moisture lapse rate at all levels and all surface temperatures between 280 and 310 K.
    \item In warm climates, increases in $e^*/p^2$ are largest in the upper troposphere. Simultaneously, $(T_0/T)\phi - 1$ decreases below 1 near the surface. Together, these changes produce large increases in the moisture lapse rate away from the surface while limiting increases close to the surface. As a result, increases in the moisture lapse rate with warming are amplified in the upper troposphere in warm climates.
\end{enumerate}

\appendix[B]
\appendixtitle{Small-domain RCE simulations}
\label{app:small_domain_rce}

We used several sets of small-domain simulations to test the robustness of two results:
\begin{enumerate}
\item Pressure velocity profile shapes collapse in $\tilde{q}$ coordinates, which allowed us to simplify the thermodynamic and dynamic modes.
\item  Height-coordinate vertical velocities at the peaks of conditional-average pressure velocity profiles scale with 99.99th percentile vertical velocities at the same level, which allowed us to connect the dynamic mode to theory for changes in peak vertical velocities.
\end{enumerate}
We found that the $\omega$ profile collapse was robust across a suite of small-domain simulations (subsection a), but that height-coordinate vertical velocities at the peaks of conditional-average pressure velocity profiles increased less quickly than 99.99th percentile vertical velocities in small-domain simulations (subsection b).

\subsection{Small-domain simulation setup}

Our small-domain simulations span the same SST range as the channel simulations and use a 128x128 km$^2$ domain with 1 km horizontal resolution. The model configuration follows the RCEMIP protocol \citep{Wing2018Radiative-convectiveProject} as closely as possible; the primary difference is that we extend the initial specific humidity profiles to our SST range by diagnosing an near-surface specific humidity that gives a relative humidity of exactly 80 percent. We ran the simulations for 100 days, allowed 60 days for the model to equilibrate, collected statistics from instantaneous snapshots taken every 6 hours over the last 40 days of simulation, and calculated condensation integrals from those snapshots in the same way as for channel simulations.

To test robustness to model configuration, we ran four sets of simulations with different combinations of microphysics and sub-grid-scale turbulence schemes. Like the channel simulation, the default small-domain simulations used the SAM single-moment microphysics scheme \citep{Khairoutdinov2003CloudSensitivities} and a first-order Smagorinsky scheme. In addition, we ran small-domain simulations with the single-moment microphysics scheme replaced by the double-moment Morrison scheme \citep{Morrison2005ADescription}, with the Smagorinsky scheme replaced by a 1.5-order turbulence closure that treats sub-grid-scale kinetic energy prognostically rather than diagnostically, and with both the single-moment microphyics and first-order turbulence schemes replaced by the double-moment microphysics and 1.5-order turbulence schemes. Like the channel simulations, all small-domain simulations calculate radiative heating rates interactively using the radiation code from the National Center for Atmospheric Research (NCAR) Community Atmosphere Model version 3 \citep[CAM,][]{Collins2006TheCAM3}.

\subsection{Robustness of the pressure velocity profile collapse}
\label{app:rce_collapse}

The collapse of conditional-average pressure velocity profiles in $\tilde{q}$ coordinates is fairly robust across all four sets of small-domain simulations (Figure A2): as in the channel simulations, pressure velocity profiles consistently peak at $\tilde{q} \approx 0.35$, and normalized pressure velocity profile shapes are close to climate-invariant. Some minor exceptions include pressure velocity profiles from the 280 K simulations with single-moment microphysics, which are shifted slightly downward relative to profiles from warmer simulations; and pressure velocity profiles from the 285 K, 290 K, and 295 K simulations with double-moment microphysics, which are slightly more bottom-heavy than profiles from warmer simulations.

The robustness of the collapse is supported by a ``collapse error'' metric that quantifies the degree to which profiles collapse in a given coordinate. The error metric is defined for two profiles $\omega_1$ and $\omega_2$ and a vertical coordinate $c$ as
\begin{equation}
    E(\omega_1, \omega_2; c) = \frac{\int \left | \omega_1 - \omega_2 \right | {\rm d}c}{\left ( \left ( \int \left | \omega_1 \right | {\rm d}c \right ) \left ( \int \left | \omega_2 \right | {\rm d}c \right )\right )^{1/2}}.
\end{equation}
In all five sets of simulations, mean collapse errors are around 0.5 when measured in pressure coordinates and are smaller by about a factor of 5 when measure in $\tilde{q}$ coordinates (Table A1), consistent with the visual impression that pressure velocity profile shapes are much closer to climate-invariant in $\tilde{q}$ coordinates than in pressure coordinates.

\subsection{Vertical velocities in small-domain RCE}
\label{app:small_domain_w}

In the channel simulations, we found that changes in height-coordinate vertical velocities at the peaks of conditional-average pressure velocity profiles closely followed changes in 99.99th percentile vertical velocities at the same level (Figure \ref{fig:f8}b). This allowed us to construct a simple model for the dynamic mode based on an entraining plume model originally formulated to understand controls on high-percentile vertical velocities in RCE. In the small-domain simulations, however, height-coordinate vertical velocities at the peaks of conditional-average pressure velocity profiles increase much less with warming than 99.99th percentile vertical velocities (Figure A3). This alters the dynamic mode relative to channel simulations and, between many sets of simulations, produces a -1 \% K$^{-1}$ to -2 \% K$^{-1}$ dynamic mode rather than the ~+2 \% K$^{-1}$ seen in channel simulations.

We do not have a complete explanation for the dynamic mode in the small-domain simulations, but we can offer some speculation as to why vertical velocities at the peaks of conditional-average profiles increase less quickly relative to vertical velocity extremes than we found in channel simulations. Conditional-average pressure velocity profiles are constructed by averaging over many individual ``member'' profiles. Plotting individual members of conditional-average profiles suggests that typical member profiles have strong upward motion over a much larger depth in channel simulations than in small-domain simulations. The presence of deep upward motion in member profiles lessens the chance that any individual member profile will have near-zero vertical velocity at the level where conditional-average profiles peak. This results in noticeable differences in the structure of individual member profiles relative to conditional-average profiles in channel versus small-domain simulations (Figure A4). At 280 K SST in both channel and small-domain simulations, very few member profiles have weak vertical motion at levels close to the peak of the conditional-average pressure velocity profiles. At 305 K SST, the channel simulation still has relatively few member profiles with weak vertical motion at the level where the conditional-average profile peaks. However, there is somewhat less coherence between different member profiles in 305 K small-domain simulations, and many member profiles show weak vertical motion at the level where the conditional-average profile peaks. 

Differences in coherence in the 305 K simulations are most apparent at the level where conditional-average profiles peak: the gray shading indicating the density of member profiles is lighter at small $\omega$ and darker at large $\omega$ in the 305 K channel simulation than in the 305 K small-domain simulation. Additionally, these differences can be quantified by calculating a mean collapse error between each member profile and the conditional-average profile as
\begin{equation}
\frac{1}{N} \sum_{i = 1}^{N} E(\omega_i, \omega_{99.99}; p),
\end{equation}
where $\omega_i$ are the $N$ member profiles and $\omega_{99.99}$ is the conditional-average profile. The mean collapse error is larger and increases more quickly with warming in small-domain simulations compared to channel simulations (Figure A5), consistent with the lack of coherence evident in Figure A4.

We can use these profiles to construct a heuristic argument for why vertical velocities at the peaks of conditional-average profiles increase less quickly than 99.99th percentile vertical velocities in small-domain simulations:

\begin{enumerate}
\item Differences between individual member profiles and conditional-average profiles are larger and increase more quickly with warming in small-domain simulations.
\item Larger differences between member profiles mean that velocities at or near 0 are sampled more often at warm temperatures from the levels where conditional-average profiles peak.
\item Sampling vertical velocities significantly above the 99.99th percentile is uncommon; more specifically, the vertical velocity sampled from member profiles only very rarely exceeds twice the 99.99th percentile vertical velocity (compare peaks of member profiles with circles in Figure A4).
\item This makes it difficult to compensate for frequently sampling vertical velocities much less than the 99.99th percentile by also frequently sampling vertical velocities much larger than the 99.99th percentile. As a result, the sample mean (i.e. the vertical velocity at the peak of the conditional-average profile) cannot increase as quickly as the 99.99th percentile (compare peaks of conditional-average profiles with circles in Figure A4).
\end{enumerate}

In other words, we speculate that differences in the relationship between 99.99th percentile vertical velocities and vertical velocities at the peak of conditional-average profiles in channel versus small-domain simulations are linked to differences in the morphology of individual updrafts (which, in turn, may be linked to differences between organized and unorganized convection). In small-domain simulations, entraining plume models remain capable of reproducing changes in 99.99th percentile vertical velocities (not shown). Because changes in vertical velocities at peaks of conditional-average profiles no longer follow changes in 99.99th percentile vertical velocities, however, we should not expect entraining plume models to reproduce the small-domain dynamic mode.

\bibliographystyle{ametsoc2014}
\bibliography{references}


\renewcommand{\thefigure}{\arabic{figure}}
\renewcommand{\thetable}{\arabic{table}}
\setcounter{figure}{0}
\setcounter{table}{0}

\begin{figure*}
\begin{center}
\includegraphics[width=38pc]{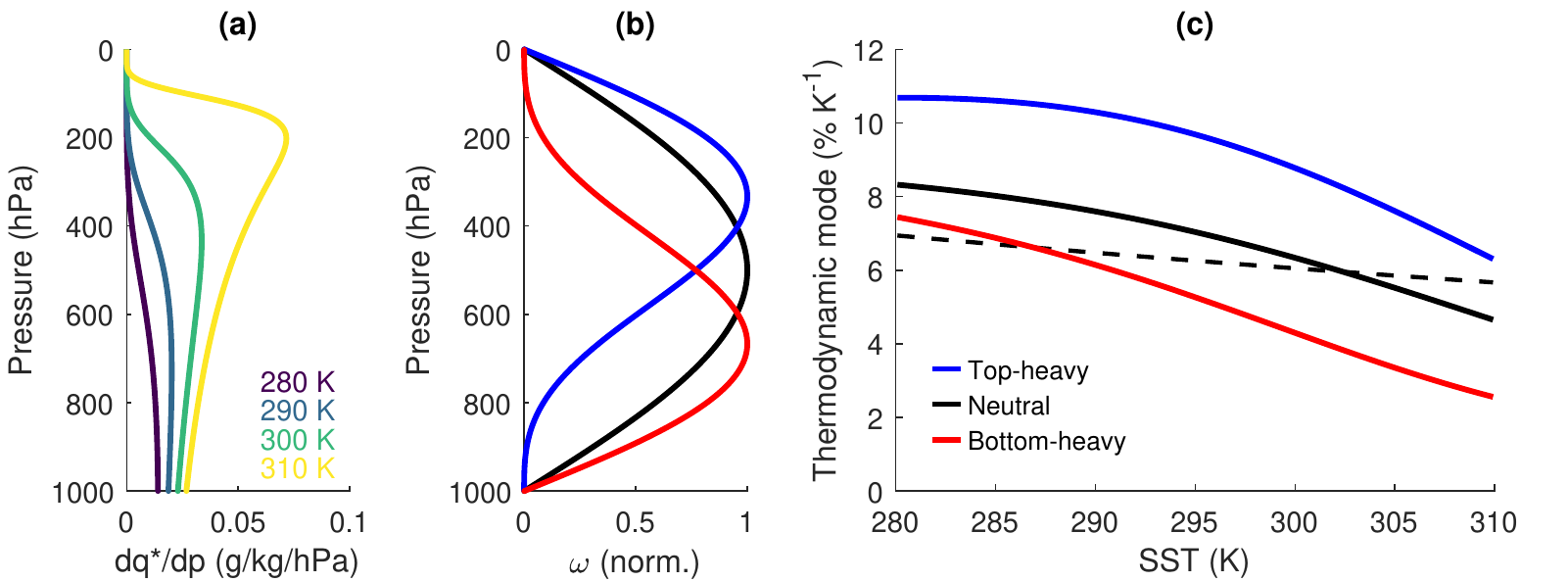}
\caption{(a) Moisture lapse rate profiles for moist adiabatic atmospheres with four surface temperatures. (b) Top-heavy (blue), neutral (black), and bottom-heavy (red) pressure velocity profile shapes. (c) Thermodynamic modes as a function of surface temperature for top-heavy, neutral, and bottom-heavy pressure velocity profiles. The dashed black line indicates a CC scaling with surface temperature.}
\label{fig:f1}
\end{center}
\end{figure*}

\begin{figure*}
\begin{center}
\includegraphics[width=38pc]{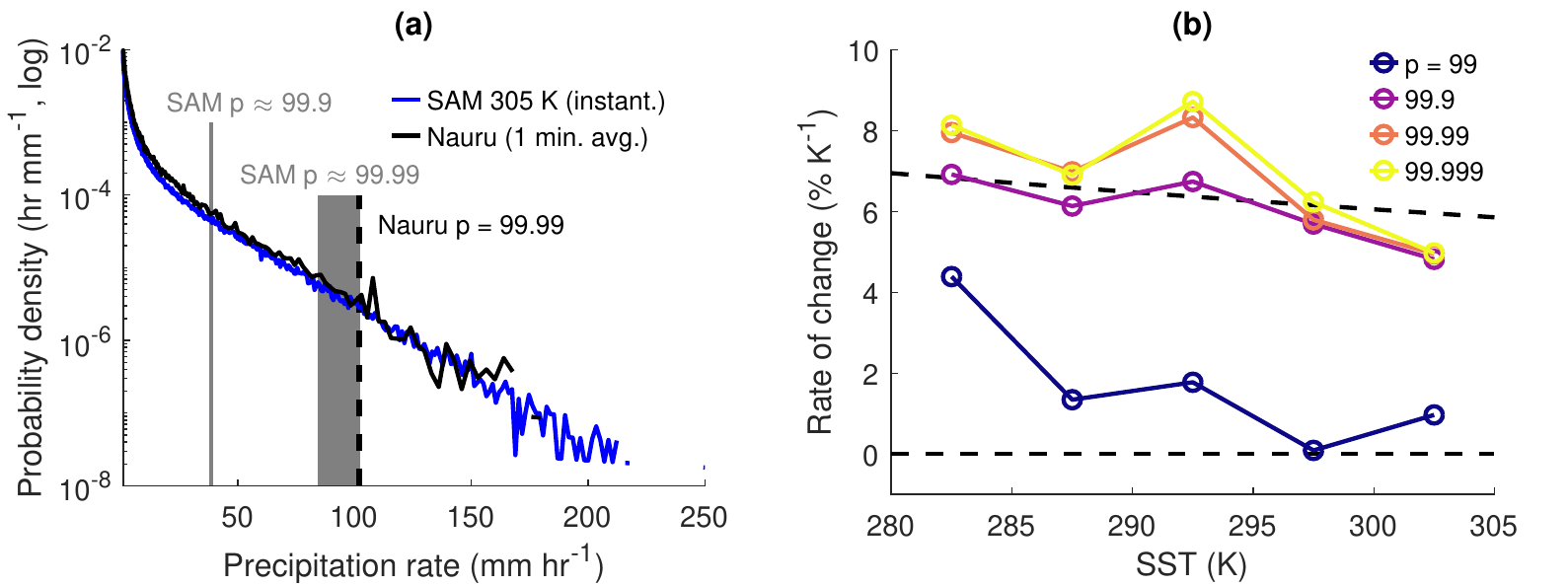}
\caption{{(a) Precipitation rate PDFs from the SAM simulation with 305 K SST (blue) and from the Nauru optical rain gauge (black). Gray shading indicates the range of events included in analyses of 99.99th percentile precipitation rates for the 305 K SAM simulation, the gray line shows the 99.9th percentile precipitation rate from the 305 K SAM simulation, and the dashed black line shows the 99.99th percentile precipitation rate from the Nauru optical rain gauge. (b) Changes in 99th, 99.9th, 99.99th, and 99.999th percentile precipitation rates with warming in SAM simulations.}}
\label{fig:f2}
\end{center}
\end{figure*}

\begin{figure*}
\begin{center}
\includegraphics[width=38pc]{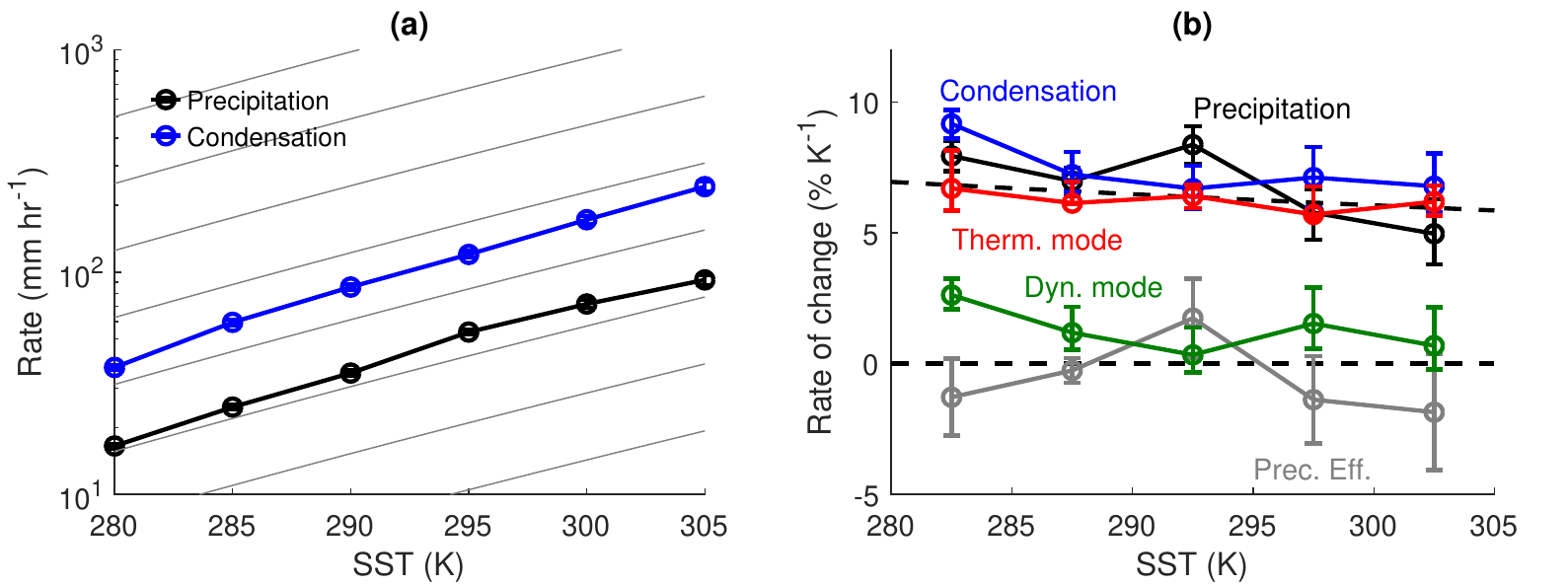}
\caption{(a) 99.99th percentile precipitation and column condensation in each SAM simulation. Gray contours slope upward following a CC scaling with SST. (b) Changes in 99.99th percentile instantaneous hydrologic extremes in SAM simulations. Black lines indicate changes in precipitation, blue lines indicate changes in column-integrated condensation, and red and green lines indicate thermodynamic and dynamic contributions to changes in column-integrated condensation. The gray line shows the contribution to changes in precipitation extremes from changes in precipitation efficiency. Black dashed lines show a CC scaling with surface temperature and zero rate of change. Error bars show sampling uncertainty estimates calculated by repeating the analysis 1000 times after resampling (with replacement) all snapshots saved during the final 25 days of the simulations.}
\label{fig:f3}
\end{center}
\end{figure*}

\begin{figure*}
\begin{center}
\includegraphics[width=38pc]{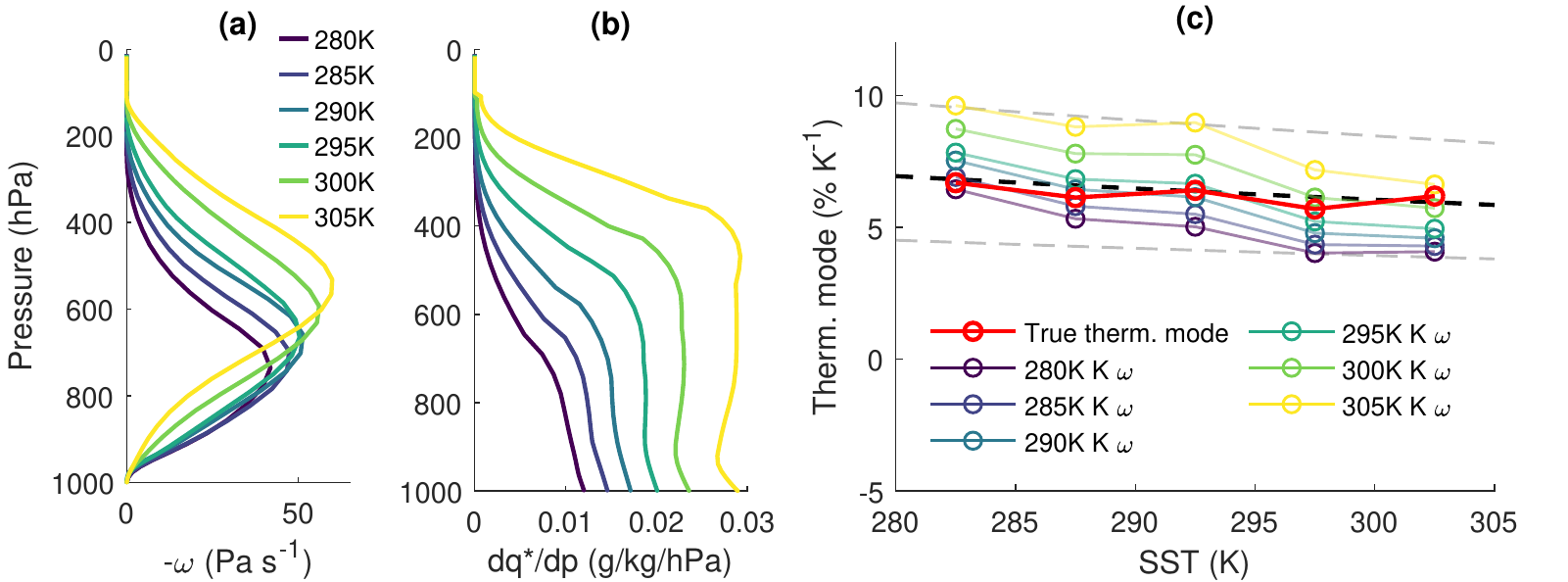}
\caption{(a) Simulated conditional-average pressure velocity profiles. (b) Moisture lapse rate profiles for each simulation. (c) Thermodynamic modes calculated while using a fixed pressure velocity profile (purple through yellow colors) compared with the true thermodynamic mode (red). The dashed black line shows a CC scaling with SST, and the dashed gray lines show 1.4 times and 0.65 times the CC scaling. The y-axis limits are identical to Figure \ref{fig:f3}.}
\label{fig:f4}
\end{center}
\end{figure*}

\begin{figure*}
\begin{center}
\includegraphics[width=38pc]{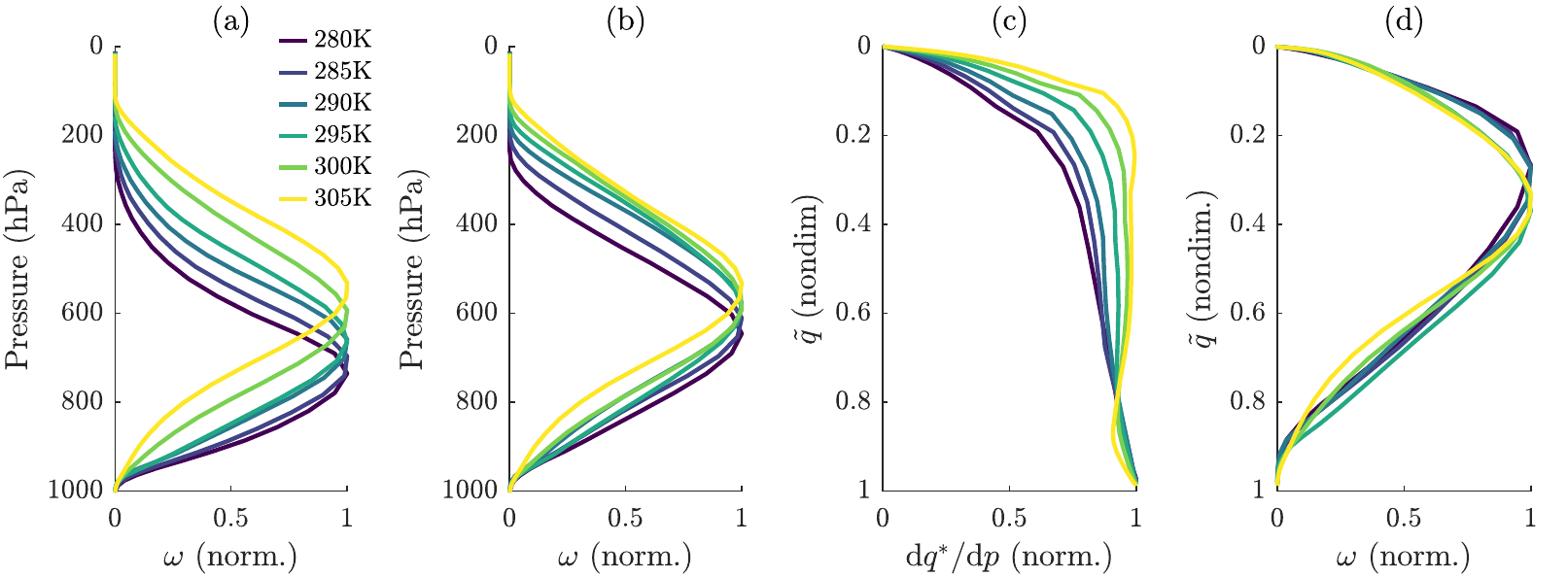}
\caption{(a,b) Simulated conditional-average pressure velocity profiles from columns with 99.99th percentile condensation rates (a) and columns with 99.99th percentile column-integrated mass fluxes (b), both plotted in pressure coordinates. (c) Moisture lapse rates plotted in $\tilde{q}$ coordinates. (d) Conditional-average pressure velocity profiles from columns with 99.99th percentile condensation rates plotted in $\tilde{q}$ coordinates.}
\label{fig:f5}
\end{center}
\end{figure*}

\begin{figure*}
\begin{center}
\includegraphics[width=19pc]{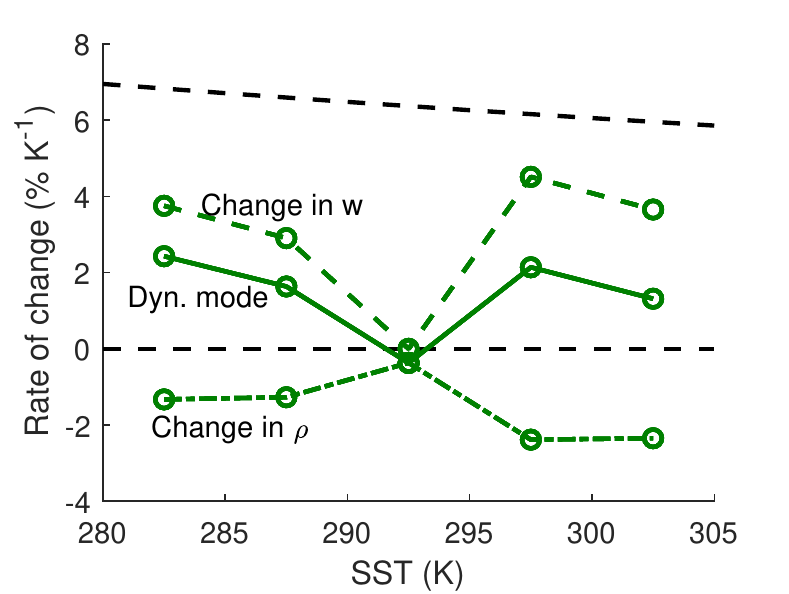}
\caption{The simplified dynamical mode (solid green line) decomposed into change in height coordinate vertical velocities (dashed green line) and densities (dash-dotted green line). Black dashed lines show a CC scaling with surface temperature and zero rate of change.}
\label{fig:f6}
\end{center}
\end{figure*}

\begin{figure*}
\begin{center}
\includegraphics[width=38pc]{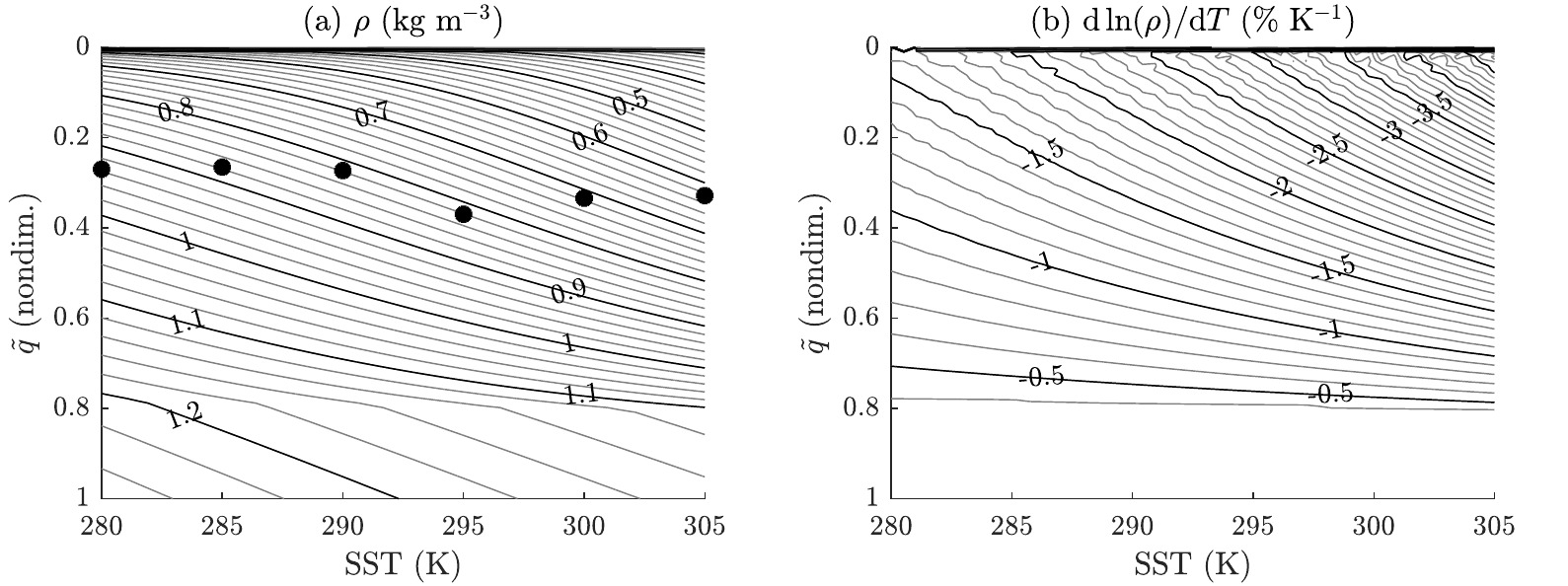}
\caption{Density (a) and rate of change in density with surface temperature (b) for a moist adiabatic atmosphere with 80\% surface relative humidity. On panel a, black dots show the levels where conditional-average pressure velocity profiles peak. Major and minor contours are spaced by 0.1 kg m$^{-3}$ and 0.02 kg m$^{-3}$ on panel a and by 0.5 \% K$^{-1}$ and 0.1 \% K$^{-1}$ on panel b.}
\label{fig:f7}
\end{center}
\end{figure*}

\begin{figure*}
\begin{center}
\includegraphics[width=38pc]{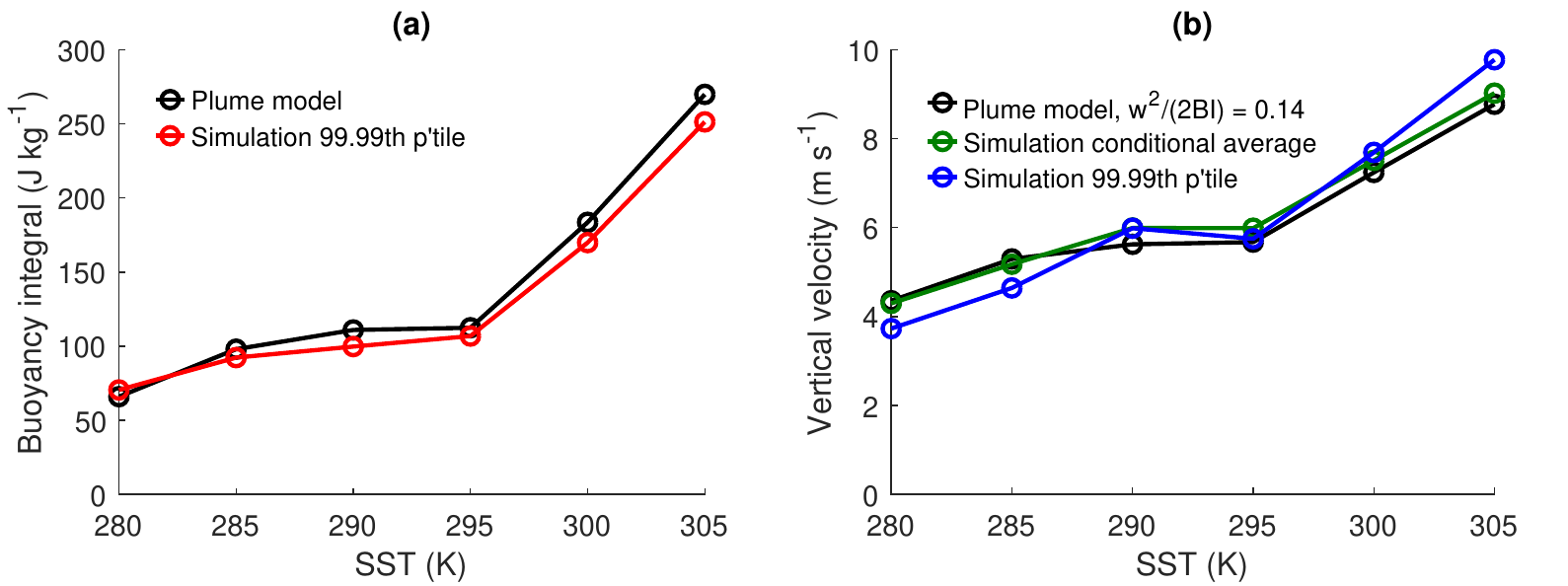}
\caption{(a) Buoyancy integrals from 1 km to the level where conditional-average pressure velocity profiles peak. The black curve shows integrals over buoyancy profiles calculated with the entraining plume model (details in text), and the red curve shows integrals over a profile constructed from level-by-level 99.99th percentile buoyancies from simulations (details in text). (b) Height-coordinate vertical velocities at the level where conditional-average pressure velocity profiles peak. The black curve shows the estimate from the entraining plume model (assuming $w^2/2$ scales with the plume model buoyancy integral with a constant of proportionality chosen empirically to be 0.14), the green curve shows the height coordinate vertical velocity from the peak of the conditional-average pressure velocity profiles, and the blue curve shows the 99.99th percentile vertical velocity at that level.}
\label{fig:f8}
\end{center}
\end{figure*}

\begin{figure*}
\begin{center}
\includegraphics[width=38pc]{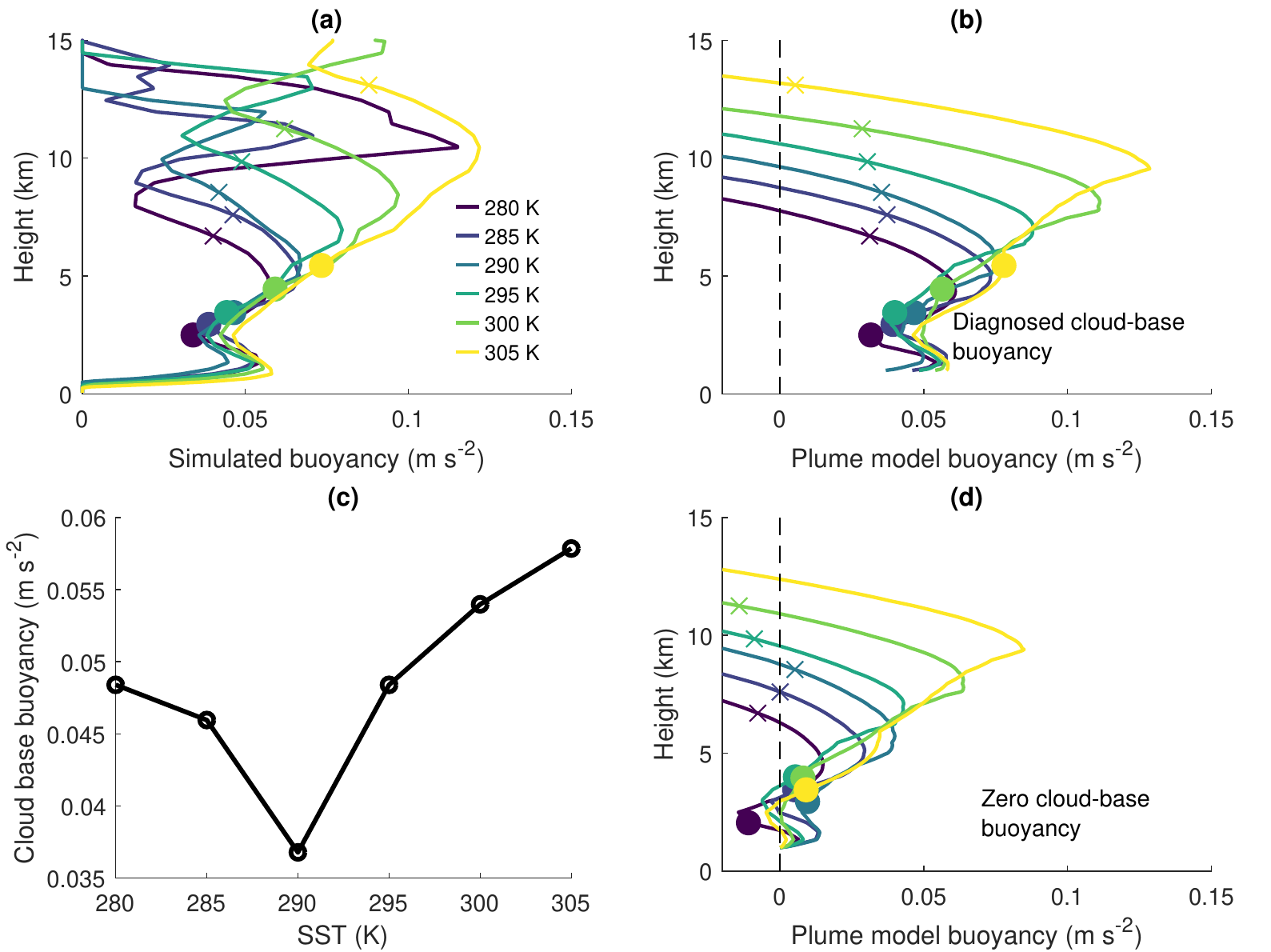}
\caption{(a,b) Buoyancy profiles from simulations (a) and plume model calculations (b) used to calculate the buoyancy integrals shown in Figure \ref{fig:f8}; (c) diagnosed cloud-base buoyancy used in the plume model calculations; (d) buoyancy profiles from alternative plume model calculations with cloud-base buoyancy set to 0. Dots on buoyancy profiles show the levels where conditional-average pressure velocity profiles peak, and x's show the level where average temperatures first drop below 220 K, approximately the tropopause temperature.}
\label{fig:f9}
\end{center}
\end{figure*}

\begin{figure*}
\begin{center}
\includegraphics[width=38pc]{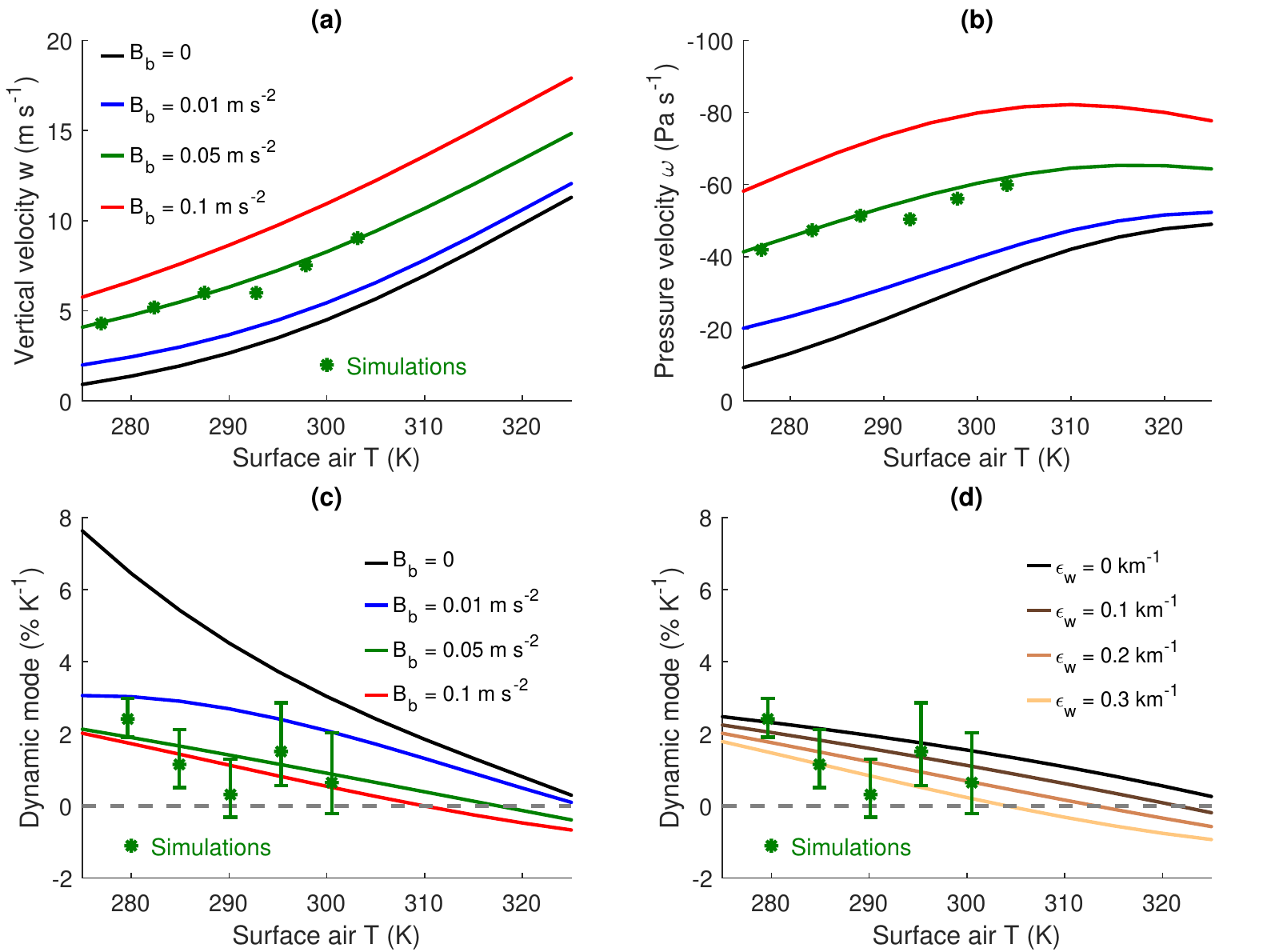}
\caption{(a-c) Height coordinate vertical velocities (a) and pressure velocities (b) at $\tilde{q} = 0.35$ from the simple model for the dynamic mode and resulting dynamic modes (c) for four values of a climate-invariant cloud-base buoyancy anomaly $B_b$. (d) The dynamic mode from the simple model with a cloud-base buoyancy anomaly of 0.05 m s$^{-2}$ and four values of the plume entrainment rate $\epsilon_w$. Green dots in panels (a) and (b) show simulated vertical velocities and pressure velocities from the peak of conditional-average pressure velocity profiles. Green dots and errorbars in panels (c) and (d) show the dynamic mode reproduced from Figure \ref{fig:f3}, but calculated and plotted using average air temperature from the lowest model level rather than SST.}
\label{fig:f11}
\end{center}
\end{figure*}

\renewcommand{\thefigure}{A\arabic{figure}}
\renewcommand{\thetable}{A\arabic{table}}
\setcounter{figure}{0}
\setcounter{table}{0}

\clearpage

\begin{figure*}
\begin{center}
\includegraphics[width=38pc]{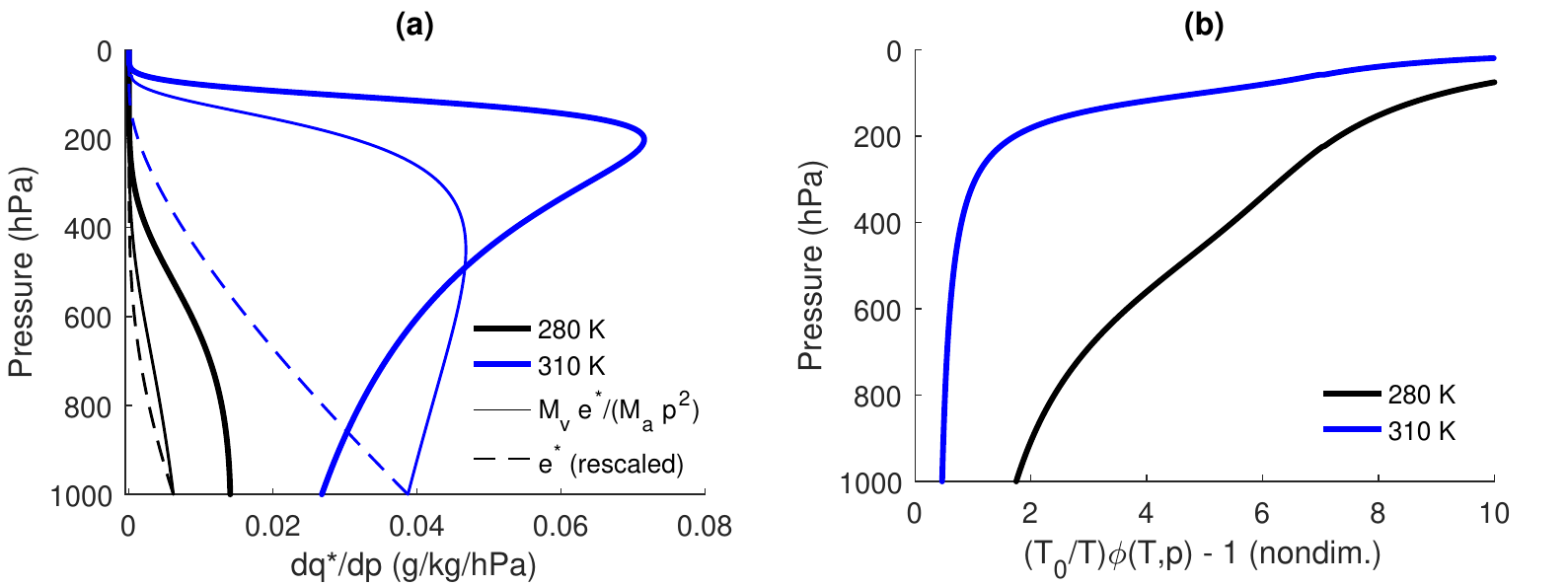}
\end{center}
\appendcaption{A1}{Moisture lapse rate profiles (a) and profiles of $(T_0/T) \phi - 1$ (b) in moist adiabatic atmospheres with two different surface temperatures. In panel (a), thin solid lines show $M_v e^*/(M_a p^2)$, and thin dashed lines show $e^*$ rescaled to have the same value as $M_v e^*/(M_a p^2)$ at the surface. See Equation \ref{eqn:mlr} for details.}
\label{fig:A0}
\end{figure*}

\begin{table*}
\label{tab:A1}
\begin{center}
\begin{tabular}{|l|c|c|c|c|c|}
\hline
& \multicolumn{1}{c|}{Channel (12288$\times$192 km$^2$)} & \multicolumn{4}{c|}{Small-domain (128$\times$128 km$^2$)} \\
\hline
& Default & Default & M2005 & 1.5-TKE & M2005 + 1.5-TKE \\
\hline
$p$ & 0.42 (0.91, 0.11) & 0.52 (1.17, 0.16) & 0.52 (1.05, 0.18) & 0.56 (1.20, 0.19) & 0.50 (1.01, 0.15) \\
$\tilde{q}$ & 0.08 (0.12, 0.02) & 0.11 (0.22, 0.04) & 0.11 (0.18, 0.04) & 0.11 (0.23, 0.03) & 0.11 (0.17, 0.02) \\
\hline
\end{tabular}
\end{center}
\appendcaption{A1}{Collapse metrics for conditional-average pressure velocity profiles in the channel simulations (see Figure \ref{fig:f5}) and small-domain simulations (see Figure A2) and in pressure coordinates (first row) and $\tilde{q}$ coordinates (second row). The first value in each cell is the mean collapse error over all pairs of profiles, and the second and third values (in parentheses) are the maximum and minimum collapse errors between any two pairs of profiles. The ``Default'' configuration refers to simulations with the SAM single-moment microphysics scheme and first-order Smagorinsky turbulence closure, ``M2005'' denotes simulations using the Morrison double-moment microphysics scheme, and ``1.5-TKE'' denotes simulations using a 1.5-order turbulence closure.}
\end{table*}

\begin{figure*}
\begin{center}
\includegraphics[width=19pc]{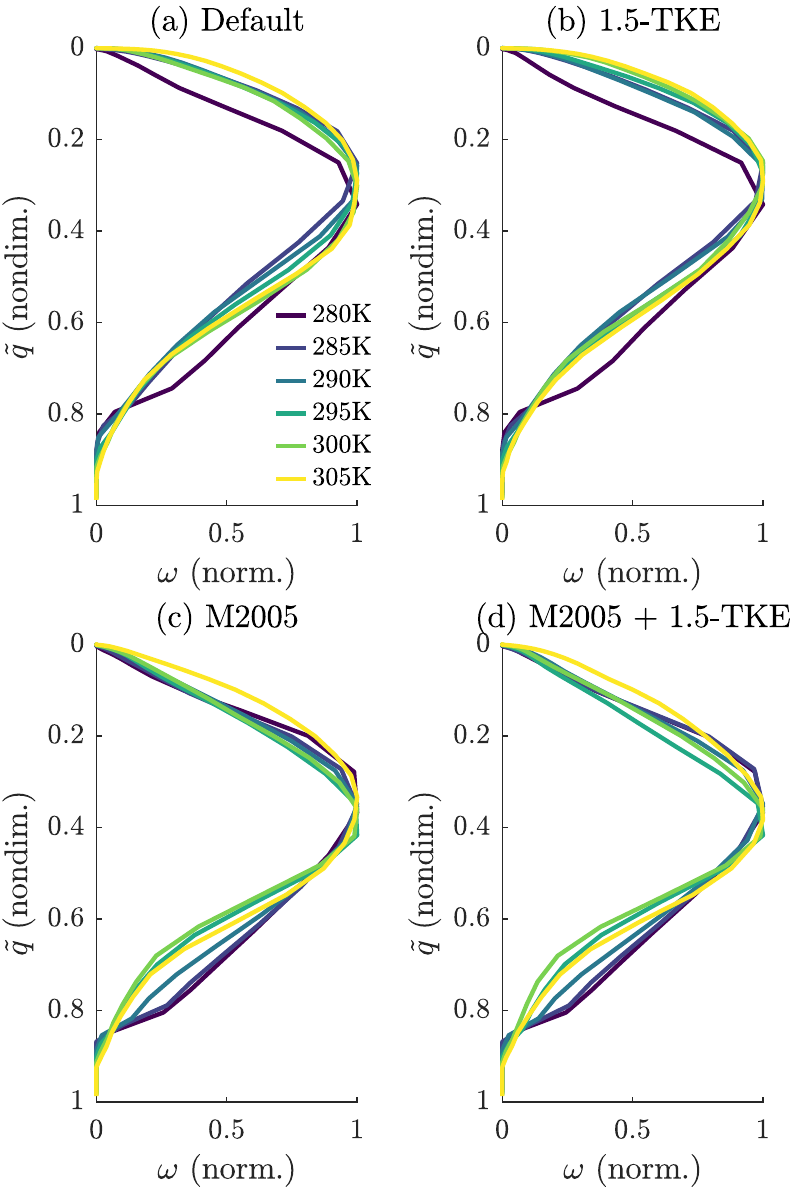}
\end{center}
\appendcaption{A2}{Conditional-average pressure velocity profiles in small-domain simulations with the SAM single-moment microphysics scheme and first-order Smagorinsky sub-grid-scale turbulence closure (a), with the 1.5-order turbulence closure (b), the Morrison double-moment microphysics scheme (c), and both the 1.5 order turbulence closure and the Morrison microphysics scheme (d).}
\label{fig:A1}
\end{figure*}

\begin{figure*}
\includegraphics[width=38pc]{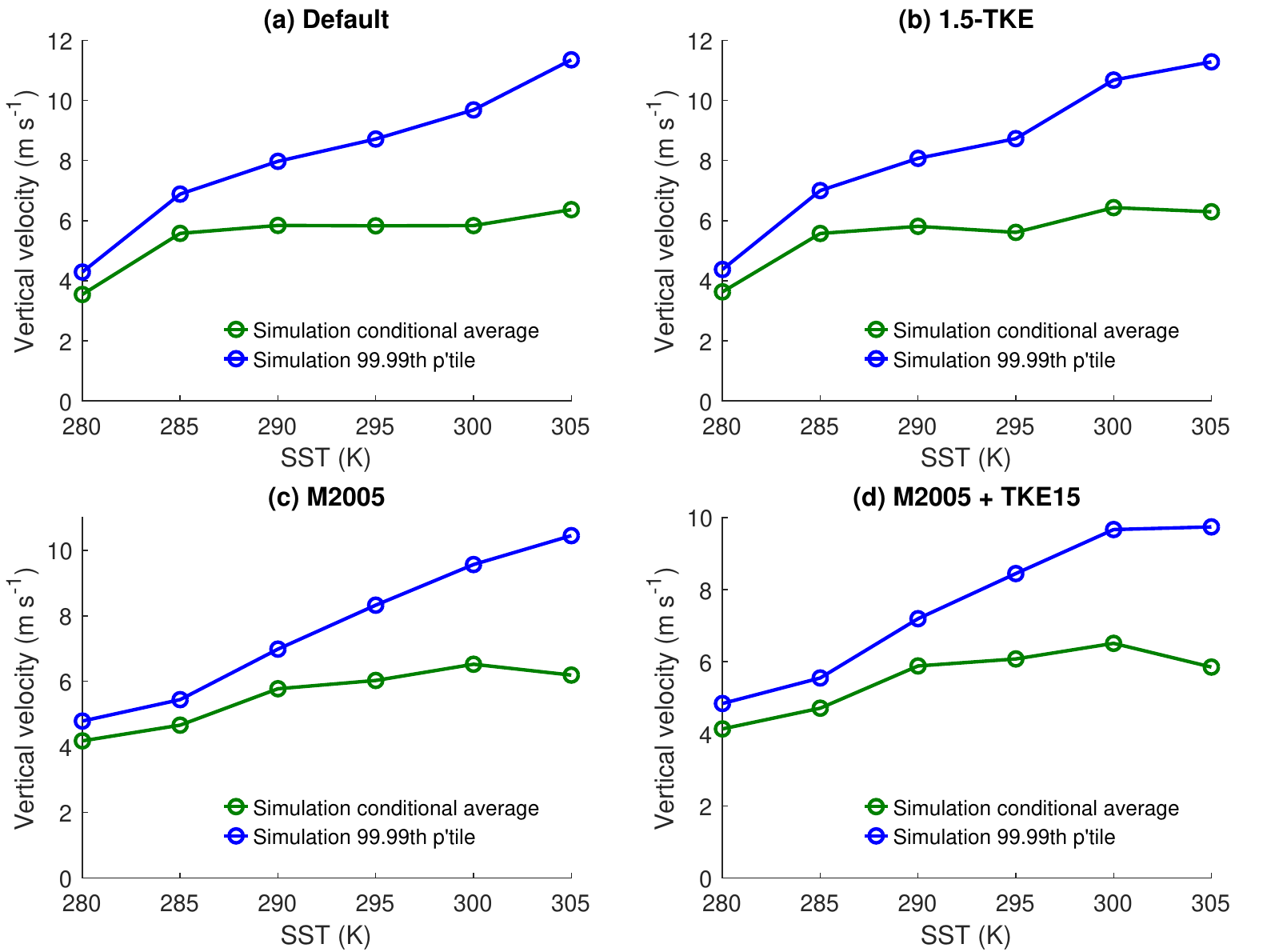}
\appendcaption{A3}{Height-coordinate vertical velocities at the level where conditional-average pressure velocity profiles peak in small-domain simulations with the SAM single-moment microphysics scheme and first-order Smagorinsky sub-grid-scale turbulence closure (a), with the 1.5-order turbulence closure (b), the Morrison double-moment microphysics scheme (c), and both the 1.5 order turbulence closure and the Morrison microphysics scheme (d). Green curves show height coordinate vertical velocities from the peaks of conditional-average pressure velocity profiles and blue curves show 99.99th percentile vertical velocities at that level.}
\label{fig:A2}
\end{figure*}

\begin{figure*}
\includegraphics[width=38pc]{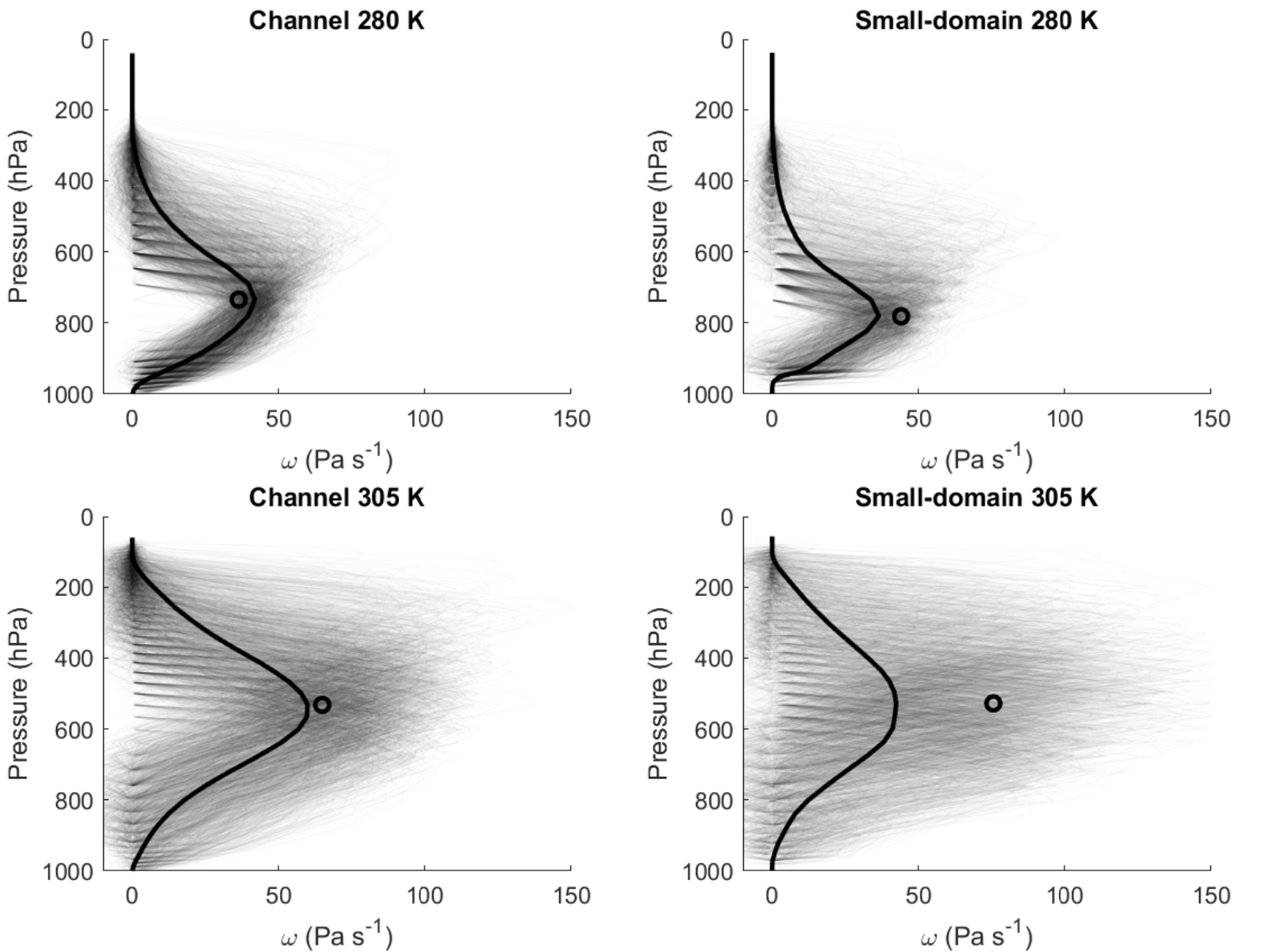}
\appendcaption{A4}{Conditional-average pressure velocity profiles (thick black lines) superimposed on the individual profiles averaged over to compute them (transluscent gray lines) from channel and small-domain simulations at 280 K and 305 K. Black circles show $\omega$ calculated from 99.99th percentile vertical velocities $w_{99.99}$ ($\omega = \rho g w_{99.99}$) at the level where conditional-average $\omega$ profiles peak.}
\label{fig:A4}
\end{figure*}

\begin{figure*}
\begin{center}
\includegraphics[width=19pc]{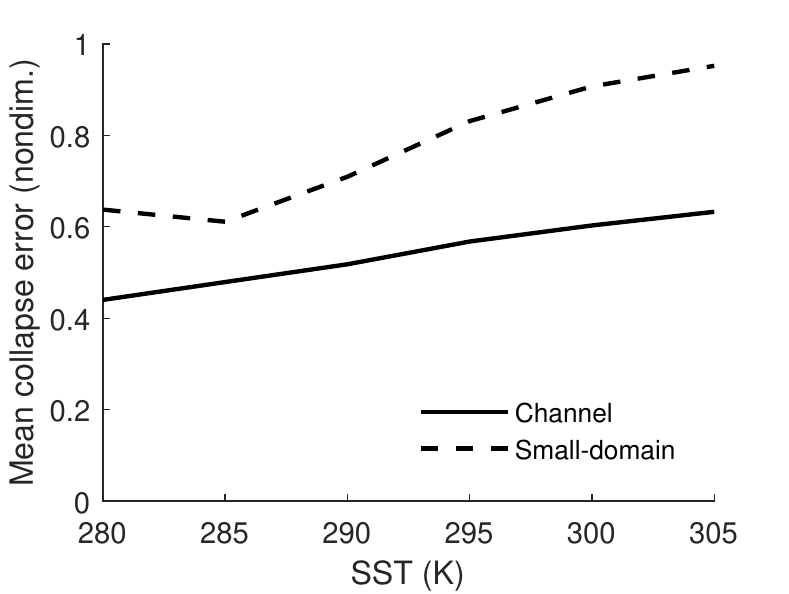}
\end{center}
\appendcaption{A5}{Average differences between individual and conditional-average $\omega$ profiles in columns with 99.99th percentile condensation, as measured by the mean collapse metric in Equation B2. The solid line shows channel simulations, the dashed line shows small-domain simulations, and smaller mean collapse errors indicate greater coherence between individual and conditional-average profiles.}
\label{fig:A5}
\end{figure*}

\end{document}